\documentclass{aa}  
\usepackage{graphicx}

\usepackage{natbib, twoopt}
\usepackage[varg]{txfonts}
\usepackage{multirow}
\usepackage{longtable}
\usepackage{lscape}

\def\sii{\relax \ifmmode {\mbox S\,{\scshape ii}}\else S\,{\scshape ii}\fi}
\def\nii{\relax \ifmmode {\mbox N\,{\scshape ii}}\else N\,{\scshape ii}\fi}
\def\oiii{\relax \ifmmode {\mbox O\,{\scshape iii}}\else O\,{\scshape iii}\fi}
\def\oi{\relax \ifmmode {\mbox O\,{\scshape i}}\else O\,{\scshape i}\fi}
\def\hii{\relax \ifmmode {\mbox H\,{\scshape ii}}\else H\,{\scshape ii}\fi}
\def\mgi{\relax \ifmmode {\mbox Mg\,{\scshape i}}\else Mg\,{\scshape i}\fi}
\def\caii{\relax \ifmmode {\mbox Ca\,{\scshape ii}}\else Ca\,{\scshape ii}\fi}

\begin{document} 

\title{Observational hints for radial migration in disc galaxies from CALIFA \thanks{Based on observations collected at the Centro Astron\'omico Hispano Alem\'an (CAHA) at Calar Alto, operated jointly by the Max-Planck Institut für Astronomie and the Instituto de Astrof\'isica de Andaluc\'ia (CSIC).}} 
\titlerunning{Observational hints for radial migration in disc galaxies from CALIFA}

\author{T. Ruiz-Lara \inst{1, 2, 3, 4}, I. P\'erez \inst{1, 2}, E. Florido \inst{1, 2},  P. S\'anchez-Bl\'azquez \inst{5}, J. M\'endez-Abreu \inst{3, 4, 6}, L. S\'anchez-Menguiano \inst{7, 1}, S. F. S\'anchez,\inst{8} M. Lyubenova \inst{9}, J. Falc\'on-Barroso \inst{3, 4}, G. van de Ven \inst{10}, R. A. Marino \inst{11}, A. de Lorenzo-Cáceres \inst{8}, C. Catal\'an-Torrecilla \inst{12}, L. Costantin \inst{13}, J. Bland-Hawthorn \inst{14}, L. Galbany \inst{15}, R. Garc\'ia-Benito \inst{7}, B. Husemann \inst{10}, C. Kehrig \inst{7}, I. M\'arquez \inst{7}, D. Mast \inst{16, 17}, C.J. Walcher \inst{18}, S. Zibetti \inst{19}, B. Ziegler \inst{20} and the CALIFA team}

\authorrunning{T. Ruiz-Lara et al.}

\institute{\inst{1} Departamento de F\'isica Te\'orica y del Cosmos, Universidad de Granada, Campus de Fuentenueva, E-18071 Granada, Spain \\
   \email{tomasruizlara@gmail.com} \\
\inst{2} Instituto Carlos I de F\'isica Te\'orica y Computacional, Universidad de Granada, E-18071 Granada, Spain \\ 
\inst{3} Instituto de Astrof\'isica de Canarias, Calle V\'ia L\'actea s/n, E-38205 La Laguna, Tenerife, Spain \\
\inst{4} Departamento de Astrof\'isica, Universidad de La Laguna, E-38200 La Laguna, Tenerife, Spain \\
\inst{5} Departamento de F\'isica Te\'orica, Universidad Aut\'onoma de Madrid, E-28049 Cantoblanco, Spain \\ 
\inst{6} School of Physics and Astronomy, University of St Andrews, SUPA, North Haugh, KY16 9SS St Andrews, UK \\ 
\inst{7} Instituto de Astrof\'isica de Andaluc\'ia (CSIC), Glorieta de la Astronom\'ia s/n, Aptdo. 3004, E-18080 Granada, Spain \\
\inst{8} Instituto de Astronom\'ia, Universidad Nacional Aut\'onoma de M\'exico, A.P. 70-264, 04510 M\'exico D.F., Mexico\\
\inst{9} Kapteyn Astronomical Institute, University of Groningen, PO Box 800, NL-9700 AV Groningen, the Netherlands \\
\inst{10} Max-Planck-Institut f\"ur Astronomie, K\"onigstuhl 17, D-69117 Heidelberg, Germany\\
\inst{11} Department of Physics, Institute for Astronomy, ETH Z\"urich, CH-8093 Z\"urich, Switzerland \\
\inst{12} Departamento de Astrof\'isica y CC. de la Atm\'osfera, Universidad Complutense de Madrid, E-28040 Madrid, Spain  \\
\inst{13} Dipartimento di Fisica e Astronomia G. Galilei, Università di Padova, vicolo dell’Osservatorio 3, I-35122 Padova, Italy   \\
\inst{14} Sydney Institute for Astronomy, School of Physics A28, University of Sydney, NSW 2006, Australia \\
\inst{15} PITT PACC, Department of Physics and Astronomy, University of Pittsburgh, Pittsburgh, PA 15260, USA \\
\inst{16} Observatorio Astron\'omico, Laprida 854, X5000BGR, C\'ordoba, Argentina  \\
\inst{17} Consejo de Investigaciones Cient\'{i}ficas y T\'ecnicas de la Rep\'ublica Argentina, Avda. Rivadavia 1917, C1033AAJ, CABA, Argentina \\
\inst{18} Leibniz-Institut f\"ur Astrophysik Potsdam (AIP), An der Sternwarte 16, D-14482 Potsdam, Germany \\
\inst{19} INAF - Osservatorio Astrofisico di Arcetri, Largo Enrico Fermi 5, I-50125 Firenze, Italy \\
\inst{20} University of Vienna, Department of Astrophysics, (T\"urkenschanzstr 17, 1180 Vienna, Austria)  \\
}

    \date{Received ---; accepted ---}


 
    \abstract
     {According to numerical simulations, stars are not always kept at their birth galactocentric distances but migrate. The importance of this radial migration in shaping galactic light distributions is still unclear. However, if it is indeed important, galaxies with different surface brightness (SB) profiles must display differences in their stellar population properties.}
     {We investigate the role of radial migration on the light distribution and the radial stellar content by comparing the inner colour, age and metallicity gradients for galaxies with different SB profiles. We define these inner parts avoiding the bulge and bar regions and up to around three disc scale-lengths (type I, pure exponential) or the break radius (type II, downbending; type III, upbending).}
     {We analyse 214 spiral galaxies from the CALIFA survey covering different SB profiles. We make use of GASP2D and SDSS data to characterise their light distribution and obtain colour profiles. The stellar age and metallicity profiles are computed using a methodology based on full-spectrum fitting techniques ({\tt pPXF}, {\tt GANDALF}, and {\tt STECKMAP}) to the IFS CALIFA data.}
     {The distributions of the colour, stellar age and stellar metallicity gradients in the inner parts for galaxies displaying different SB profiles are unalike as suggested by Kolmogorov-Smirnov and Anderson-Darling tests. We find a trend in which type II galaxies show the steepest profiles of all and type III the shallowest, with type I galaxies displaying an intermediate behaviour.}
     {These results are consistent with a scenario in which radial migration is more efficient for type III galaxies than for type I systems with type II galaxies presenting the lowest radial migration efficiency. In such scenario, radial migration mixes the stellar content flattening the radial stellar properties and shaping different SB profiles. However, in sight of these results we cannot further quantify its importance in shaping spiral galaxies, and other processes such as recent star formation or satellite accretion might play a role.} 


    \maketitle
%

\section{Introduction}
\label{II_intro}

The analysis of the stellar content in galaxies is an essential tool unveiling the processes that these systems underwent throughout their history \citep[e.g.][]{2009MNRAS.395...28M, 2011MNRAS.416.1996R, 2011MNRAS.415..709S, 2011A&A...529A..64P, 2013ApJ...764L...1P}. Stellar populations reflect the chemistry of the interstellar medium at the moment of their formation and temporal variations in the star formation activity can give us essential information about the past of the galaxy \citep[][]{2007ApJ...659L..17C, 2015ApJ...811L..18G, 2015A&A...581A.103G, 2015MNRAS.451.3400B, 2016A&A...590A..44G}. However, galaxies are dynamical systems in continuous change and stars do not remain at their birth locations. We must take into account the effect of stellar motions to properly interpret stellar population information \citep[e.g.][]{2008ApJ...675L..65R, 2009ApJ...705L.133M}.

Several theoretical works have studied why stars undergo radial motions and its effects on disc properties \citep[][]{2002MNRAS.336..785S, 2006ApJ...645..209D, 2007ApJ...670..269Y, 2008ApJ...684L..79R, 2008ApJ...675L..65R, 2009ApJ...705L.133M, 2009MNRAS.398..591S, 2010ApJ...722..112M, 2011A&A...527A.147M, 2012MNRAS.426.2089R, 2012A&A...548A.126M, 2012MNRAS.420..913B}. The causes for such motions can be attributed to internal (mainly caused by axisymmetric structures) or external (effect of satellite influence) agents. \citet[][]{2002MNRAS.336..785S} proposed that stars close to the corotation resonance of transient spirals experience large changes in their radial positions. In addition, a non-linear coupling of non-axisymmetric structures such as the bar and the spiral structure leads to stronger migrations to those caused by the single presence of transient spirals \citep[][]{2010ApJ...722..112M, 2012A&A...548A.126M, 2012A&A...548A.127M}. The influence of nearby satellites, as well as satellite accretion, can also induce mixing in the stellar discs \citep[][]{2007ApJ...670..269Y, 2012MNRAS.420..913B}.

Recent simulations suggest that these radial motions have a considerable effect on the galaxy properties. In \citet{2008ApJ...684L..79R}, using N-body and smoothed-particle hydrodynamics (SPH) simulations of an isolated and idealized disc, the authors found mass profiles with a lack of mass in the outer parts and age profiles with a characteristic ``U-shape'' (i.e. an age radial decline followed by an outer upturn). They suggested that these features found in the outer parts (i.e. lack of mass and outer ageing) are attributed to the interplay between a radial star formation cutoff and radial redistribution of stars induced by transient spiral arms. \citet[][]{2009MNRAS.398..591S}, using fully-cosmological hydrodynamical simulations, also found a downbending light surface density profile (lack of light in the outer parts) and an ``U-shape'' age profile in their simulated disc. In this case, the authors claim that breaks in the light distribution do not necessarily correspond to breaks in the mass distribution suggesting that the break origin is linked to two main processes: i) a radial change in the slope of the star formation profile linked to a drop in the gas density (due to a warp) as the main cause and ii) radial migration of stars towards larger radii. It is important to note though that such ``U-shaped'' age gradient was found even in the absence of radial migration. Although theoretical works are concentrated on galaxies displaying downbending profiles (type II), several observational works have also observed systems with pure exponential (type I) declines of the light profiles and upbending (type III) Surface Brightness (SB) distributions \citep[e.g.][]{2005ApJ...629..239B, 2006A&A...454..759P, 2008AJ....135...20E, 2011AJ....142..145G, 2016A&A...585A..47M}.

Although there is no clear consensus on the causes of radial migration or a definite explanation for the simulated SB and age profiles, all these works point towards an important amount of stars migrating from the inner regions to the outer parts. In particular, \citet[][]{2009MNRAS.398..591S} found that 57~\% of the stars currently located in the outer parts of their cosmologically-simulated disc came from the inner region, with mean values of the traversed radial distance of $\sim$ 3.4 kpc. Other works have also analysed the amount of stellar particles populating the outer parts and coming from the inner regions: \citet[][]{2008ApJ...675L..65R} found that the percentage of outwards migrating stars is up to $\sim$ 85~\% with the average change in radius being 3.7 kpc, while \citet[][]{2009ApJ...705L.133M} obtained a percentage as high as 64~\% -- 78~\% in their cosmological discs. In \citet[][]{2012MNRAS.426.2089R}, they analysed in detail the origin of the radial migration observed in their idealised and isolated discs obtaining that nearly 50~\% of the stars populating their solar neighbourhood (7 < R[kpc] < 9) came from the inner disc with some of them experiencing radial changes as high as 7 kpc (although it is not the norm). Observationally, evidence of radial migration has also been found in stars of the Milky Way \citep[][]{2001A&A...377..911F, 2004A&A...418..989N, 2014A&A...565A..89B}. In particular, the RAVE collaboration \citep[][]{2006AJ....132.1645S, 2013AJ....146..134K} has recently found that around half of the supersolar metallicity stars currently located in the solar neighbourhood have migrated from inner regions \citep[][]{2015MNRAS.447.3526K}. This radial redistribution of material should affect, not only the outer parts, but the overall stellar population and light distributions, especially if the number of migrating stars is high.

In particular, in \citet[][]{2009MNRAS.398..591S}, where the broken profiles were caused by the combination of a radial change in the star formation rate linked to a warp and radial migration, the authors speculate that the different observed SB profiles in the literature might be explained by different combinations of both processes. If we ignore the effect of the warp, we would expect a smooth change in the star formation rate per area unit leading to a pure exponential profile. Depending on how intense radial migration is, we could change from galaxies displaying a type II surface density profile (systems with little outwards radial migration) to galaxies displaying a type III profile (with a higher efficiency of the outwards radial migration) with type I galaxies being a case in between. Thus, different radial migration efficiencies might produce different SB profiles. If this scenario proposed by \citet[][]{2009MNRAS.398..591S} is correct, then, galaxies displaying different SB distributions must present differences in the stellar content throughout their discs. Therefore, the analysis of the stellar populations from the inner regions up to the outer discs of spiral galaxies is essential to better understand the role of radial migration, the general assembly of spiral galaxies, and to refine and constrain galaxy formation models. This analysis can be carried out following different approaches.

One possible approach consists of taking photometric images using different broad-band filters. As a first approximation, light differences from different filters (colours) can be interpreted as variations in the properties of the stellar content. A vast amount of works used this approach to determine stellar population gradients \citep[e.g. ][]{1993A&A...271...51P, 1996A&A...313..377D, 1996AJ....111.2238P, 2000ApJS..126..331J, 2000MNRAS.312..497B, 2004ApJS..152..175M, 2009ApJ...703.1569M, 2011MNRAS.416.1983R, 2011MNRAS.416.1996R, 2012ApJ...758...41R}. Whether optical colours can be interpreted as a proxy for stellar age or metallicity is still controversial.

A different approach to analyse the stellar light is by using spectroscopic data. This approach allows us to study specific features dependent on the stellar age and metallicity, i.e. the line--strength indices \citep[e.g.][]{1984AJ.....89.1238R, 1985ApJS...57..711F, 1986A&A...162...21B, 1986A&AS...66..171B, 1988A&A...195...76B, 1993ApJS...86..153G, 1994ApJS...95..107W, 1994A&A...283..805B, 1996ApJS..106..307V, 1997ApJS..111..377W, 2003MNRAS.341...33K}. Indices have been used to obtain single stellar population (SSP) equivalent values for age and metallicity in ``simple'' systems such as globular clusters or elliptical galaxies \citep[e.g.][]{2007MNRAS.379..445P, 2010MNRAS.408...97K}. However, in the recent years great efforts have been put to improve the quality of the information recovered from indices and to make possible the application of this approach to any kind of system \citep[e.g.][]{2005MNRAS.362...41G, 2006MNRAS.370.1106G, 2008MNRAS.383.1439G}. The combination of recently developed bayesian approaches, improved spectral libraries, and high-quality observed spectra has allowed studies based on indices to obtain star formation histories, as well as metallicity and dust distributions \citep[][]{2017MNRAS.468.1902Z}. However, the needed signal-to-noise (S/N) hampers this kind of analysis in regions of low SB or with data of limited quality.

In order to overcome this issue and to maximise the information used from observed spectra, several codes have been developed in the recent years analysing wide wavelength ranges \citep[e.g.][]{2000MNRAS.317..965H, 2001MNRAS.327..849R, 2005MNRAS.358..363C, 2007MNRAS.381.1252T, 2006MNRAS.365...46O, 2006MNRAS.365...74O, 2009A&A...501.1269K}. This full-spectrum fitting approach has been proven successful at further minimising the well-known age-metallicity degeneracy \citep[][]{2011MNRAS.415..709S}. However, studies of the stellar content on spiral galaxies are still very scarce, limited to the bulges or the inner discs \citep[][]{2009MNRAS.395...28M, 2011MNRAS.415..709S, 2011MNRAS.410..313S, 2014MNRAS.437.1534S, 2015MNRAS.452.1128M, 2016MNRAS.463.4396M}. 

The emergence of integral field spectroscopy (IFS) makes possible a reliable analysis of the stellar content up to larger galactocentric distances. The IFS instruments and data at our disposal allow us to carry out new stellar population studies with unprecedented quality \citep[][]{2012ApJ...752...97Y, 2014A&A...570A...6S, 2015A&A...581A.103G, 2016MNRAS.456L..35R}.

In this paper we study the stellar content of a sample of 214 spiral galaxies from the CALIFA survey \citep[][]{2012A&A...538A...8S} using full spectrum fitting techniques in order to asses, for the first time in such a large sample, the role of radial migration in shaping the properties that we observe in spiral galaxies. In addition, we analyse photometric data from the Sloan Digital Sky Survey \citep[SDSS,][]{2000AJ....120.1579Y} to characterise the two-dimensional light distribution of these galaxies. This analysis allows us to test if galaxies with the same SB profiles display similarities in their inner stellar content. Besides, we also obtain SDSS colour profiles ($g-r$, $g-i$, and $r-i$) for a further comparison. In Sect.~\ref{II_sample} we define the sample of galaxies under study. We explain the method to analyse SDSS and CALIFA data in Sects.~\ref{II_SDSS_gasp2d} and~\ref{II_spec_analysis}. The main results, discussion, and conclusions of this work are given in Sects.~\ref{II_results} and~\ref{II_conclusions}.


\section{Sample selection}
\label{II_sample}

The sample of galaxies under analysis in this work was chosen from the CALIFA mother sample \citep{2014A&A...569A...1W} as well as the CALIFA extension presented in the CALIFA third Data Release \citep[DR3, ][]{2016A&A...594A..36S}. We have selected a subset of disc galaxies (S0/a to Sd) with no signs of interaction according to the sample characterisation presented in \citet{2014A&A...569A...1W}. In addition, the galaxies under analysis are those analysed in \citet[][]{2017A&A...598A..32M} for which accurate two-dimensional (2D) decomposition of their light distribution is available. Galaxies fulfilling all these criteria but with inclinations larger than 70$^\circ$ or for which the 2D light decomposition does not find any disc component (those dominated by a spheroid component) are discarded.

\begin{figure*}
\centering
\includegraphics[width = 0.9\textwidth]{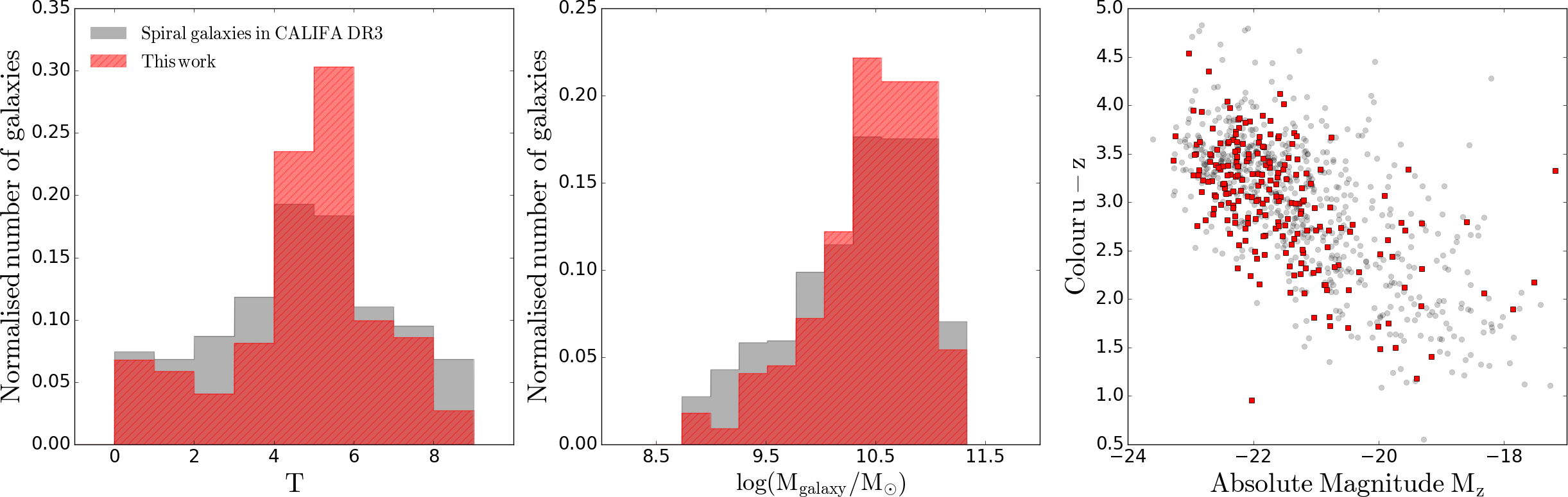} \\
\caption{Characterisation of the sample of galaxies under analysis in this work (red colour) compared to the CALIFA DR3 sample of galaxies limited to spiral systems (grey colours). Left-hand panel: Distribution of the morphological types according to the T parameter from \citet{2014A&A...569A...1W}. Middle panel: Distribution of stellar masses from \citet{2014A&A...569A...1W}. Right-hand panel: Distribution of the galaxies in the $(u-z)$ -- $M_z$ colour magnitude diagram.}
\label{II_sample_char}
\end{figure*}

The final sample comprises 214 galaxies (124 barred galaxies, 58 \%; and 90 unbarred galaxies, 42 \%). Figure\,\ref{II_sample_char} is aimed at characterising the morphological and mass distributions as well as the position in a $(u-z)$ -- $M_z$ colour magnitude diagram of the selected galaxies (red). In order to compare with the general behaviour of the CALIFA DR3 spiral galaxies, the characteristics of this set of spiral systems are also shown (grey). We can highlight that the galaxies analysed in this study are representative of the CALIFA sample with the addition of the galaxies from the extension projects (not necessarily fulfilling all the CALIFA sample criteria) and thus, unbiased to any particular mass value or morphological type. Table \ref{II_galaxy_tab} lists the main characteristics of each individual galaxy.

\begin{longtab}
\centering 
\small
\begin{longtable}{lllllllll}
\hline\hline\\
Galaxy & RA & Dec. & Morphological & Bar & SB & h$_{\rm in}$ & h$_{\rm out}$ & R$_{\rm break}$/h$_{\rm in}$ \\ 
 & (deg) & (deg) & Type & (1, yes; 0 no) & profile & (kpc) & (kpc) &  \\ 
$(1)$ & $(2)$ & $(3)$ & $(4)$ & $(5)$ & $(6)$ & $(7)$ & $(8)$ & $(9)$ \\ 
\hline\\
\endfirsthead
\hline\hline\\
Galaxy & RA & Dec. & Morphological & Bar & SB & h$_{\rm in}$ & h$_{\rm out}$ & R$_{\rm break}$/h$_{\rm in}$ \\ 
 & (deg) & (deg) & Type & (1, yes; 0 no) & profile & (kpc) & (kpc) &  \\ 
$(1)$ & $(2)$ & $(3)$ & $(4)$ & $(5)$ & $(6)$ & $(7)$ & $(8)$ & $(9)$ \\ 
\hline\\
\endhead
\caption{Continued.}\\
\endfoot
\endlastfoot
UGC~00005  & 0.77 &  -1.91  & 5.0 &  0  & 2 &  6.24 &  3.92 &  2.71 \\
NGC~7819  & 1.1 &  31.47  & 6.0 &  1  & 2 &  15.32 &  4.16 &  1.05 \\
IC~1528  & 1.27 &  -7.09  & 5.0 &  0  & 1 &  4.22 &  -- &  -- \\
UGC~00036  & 1.31 &  6.77  & 3.0 &  1  & 1 &  4.07 &  -- &  -- \\
NGC~0001  & 1.82 &  27.71  & 5.0 &  0  & 1 &  4.3 &  -- &  -- \\
NGC~0023  & 2.47 &  25.92  & 4.0 &  1  & 2 &  34.34 &  16.7 &  0.46 \\
NGC~0036  & 2.84 &  6.39  & 4.0 &  1  & 1 &  9.57 &  -- &  -- \\
UGC~00139  & 3.63 &  -0.74  & 7.0 &  0  & 3 &  3.27 &  4.75 &  2.49 \\
MCG-02-02-030  & 7.53 &  -11.11  & 4.0 &  0  & 2 &  4.51 &  2.61 &  1.17 \\
UGC~00312  & 7.85 &  8.47  & 8.0 &  1  & 2 &  5.53 &  1.19 &  1.61 \\
ESO540-G003  & 8.91 &  -20.13  & 4.0 &  1  & 1 &  3.39 &  -- &  -- \\
NGC~0160  & 9.02 &  23.96  & 2.0 &  0  & 2 &  17.02 &  5.08 &  1.03 \\
NGC~0165  & 9.12 &  -10.11  & 4.0 &  1  & 1 &  5.79 &  -- &  -- \\
NGC~0171  & 9.34 &  -19.93  & 4.0 &  1  & 2 &  20.4 &  4.1 &  0.64 \\
NGC~0180  & 9.49 &  8.64  & 4.0 &  1  & 1 &  8.43 &  -- &  -- \\
NGC~0214  & 10.37 &  25.5  & 5.0 &  1  & 1 &  4.27 &  -- &  -- \\
NGC~0237  & 10.87 &  -0.12  & 6.0 &  0  & 1 &  2.71 &  -- &  -- \\
NGC~0234  & 10.88 &  14.34  & 6.0 &  0  & 2 &  5.57 &  3.11 &  1.55 \\
NGC~0257  & 12.01 &  8.3  & 6.0 &  0  & 1 &  5.56 &  -- &  -- \\
NGC~0309  & 14.18 &  -9.91  & 7.0 &  1  & 1 &  9.27 &  -- &  -- \\
NGC~0447  & 18.91 &  33.07  & 2.0 &  1  & 1 &  10.13 &  -- &  -- \\
NGC~0477  & 20.34 &  40.49  & 5.0 &  0  & 1 &  6.75 &  -- &  -- \\
IC~1683  & 20.66 &  34.44  & 4.0 &  1  & 1 &  4.89 &  -- &  -- \\
NGC~0496  & 20.8 &  33.53  & 7.0 &  0  & 2 &  5.52 &  2.66 &  3.1 \\
NGC~0528  & 21.39 &  33.67  & 0.0 &  0  & 3 &  2.93 &  5.96 &  3.57 \\
NGC~0551  & 21.92 &  37.18  & 5.0 &  1  & 1 &  5.11 &  -- &  -- \\
NGC~0570  & 22.24 &  -0.95  & 4.0 &  1  & 1 &  5.66 &  -- &  -- \\
UGC~01271  & 27.25 &  13.21  & 1.0 &  1  & 1 &  4.02 &  -- &  -- \\
NGC~0716  & 28.25 &  12.71  & 4.0 &  1  & 1 &  3.36 &  -- &  -- \\
NGC~0768  & 29.67 &  0.53  & 6.0 &  0  & 2 &  6.55 &  2.63 &  3.12 \\
NGC~0776  & 29.98 &  23.64  & 4.0 &  1  & 1 &  5.67 &  -- &  -- \\
NGC~0787  & 30.2 &  -9.0  & 2.0 &  0  & 3 &  4.03 &  6.48 &  3.26 \\
NGC~0842  & 32.46 &  -7.76  & 0.0 &  1  & 2 &  4.33 &  3.31 &  2.41 \\
UGC~01659  & 32.49 &  16.03  & 6.0 &  1  & 2 &  29.83 &  4.74 &  0.33 \\
NGC~0873  & 34.13 &  -11.35  & 7.0 &  0  & 1 &  2.77 &  -- &  -- \\
NGC~0924  & 36.7 &  20.5  & 0.0 &  1  & 1 &  5.49 &  -- &  -- \\
UGC~01918  & 36.89 &  25.67  & 4.0 &  1  & 2 &  6.27 &  3.36 &  0.98 \\
NGC~0932  & 36.98 &  20.33  & 1.0 &  0  & 1 &  5.51 &  -- &  -- \\
NGC~0941  & 37.12 &  -1.15  & 7.0 &  0  & 1 &  2.23 &  -- &  -- \\
NGC~0976  & 38.5 &  20.98  & 5.0 &  1  & 1 &  3.3 &  -- &  -- \\
NGC~0991  & 38.89 &  -7.15  & 7.0 &  1  & 2 &  2.91 &  1.3 &  2.35 \\
UGC~02099  & 39.3 &  21.57  & 1.0 &  0  & 1 &  10.25 &  -- &  -- \\
UGC~02134  & 39.72 &  27.85  & 4.0 &  0  & 2 &  21.27 &  4.11 &  0.41 \\
NGC~1070  & 40.84 &  4.97  & 4.0 &  0  & 3 &  2.38 &  5.45 &  2.6 \\
NGC~1094  & 41.87 &  -0.29  & 4.0 &  0  & 3 &  2.72 &  5.55 &  3.17 \\
NGC~1093  & 42.07 &  34.42  & 5.0 &  1  & 1 &  4.56 &  -- &  -- \\
UGC~02311  & 42.37 &  -0.87  & 5.0 &  1  & 1 &  5.72 &  -- &  -- \\
UGC~02403  & 43.99 &  0.69  & 4.0 &  1  & 1 &  4.95 &  -- &  -- \\
UGC~02443  & 44.59 &  -2.04  & 7.0 &  0  & 2 &  3.68 &  1.39 &  1.28 \\
NGC~1167  & 45.43 &  35.21  & 0.0 &  0  & 1 &  8.71 &  -- &  -- \\
NGC~1211  & 46.72 &  -0.79  & 1.0 &  1  & 2 &  6.49 &  1.67 &  0.67 \\
UGC~02690  & 50.18 &  -1.11  & 7.0 &  0  & 2 &  4.83 &  2.57 &  3.33 \\
MCG-01-10-019  & 55.18 &  -6.42  & 5.0 &  0  & 1 &  7.1 &  -- &  -- \\
NGC~1645  & 71.03 &  -5.47  & 1.0 &  1  & 1 &  5.32 &  -- &  -- \\
NGC~1659  & 71.62 &  -4.79  & 5.0 &  1  & 1 &  3.68 &  -- &  -- \\
NGC~1665  & 72.07 &  -5.43  & 0.0 &  0  & 1 &  3.45 &  -- &  -- \\
NGC~1666  & 72.14 &  -6.57  & 1.0 &  1  & 2 &  2.55 &  1.84 &  9.14 \\
NGC~1667  & 72.15 &  -6.32  & 5.0 &  1  & 1 &  3.58 &  -- &  -- \\
UGC~03253  & 79.92 &  84.05  & 4.0 &  1  & 1 &  3.59 &  -- &  -- \\
NGC~2253  & 100.92 &  65.21  & 5.0 &  1  & 1 &  2.89 &  -- &  -- \\
NGC~2347  & 109.02 &  64.71  & 5.0 &  1  & 1 &  4.66 &  -- &  -- \\
UGC~03944  & 114.65 &  37.63  & 5.0 &  1  & 1 &  3.16 &  -- &  -- \\
UGC~03973  & 115.64 &  49.81  & 5.0 &  1  & 1 &  7.6 &  -- &  -- \\
UGC~03995  & 116.04 &  29.25  & 4.0 &  1  & 1 &  7.39 &  -- &  -- \\
NGC~2449  & 116.83 &  26.93  & 3.0 &  1  & 1 &  4.49 &  -- &  -- \\
NGC~2487  & 119.59 &  25.15  & 4.0 &  1  & 1 &  7.39 &  -- &  -- \\
UGC~04195  & 121.28 &  66.78  & 4.0 &  1  & 2 &  6.75 &  2.62 &  1.87 \\
NGC~2530  & 121.98 &  17.82  & 8.0 &  1  & 1 &  4.65 &  -- &  -- \\
NGC~2540  & 123.19 &  26.36  & 5.0 &  1  & 2 &  5.34 &  2.21 &  2.79 \\
NGC~2543  & 123.24 &  36.25  & 5.0 &  1  & 1 &  3.81 &  -- &  -- \\
UGC~04308  & 124.36 &  21.69  & 6.0 &  1  & 2 &  4.19 &  2.32 &  2.2 \\
NGC~2553  & 124.4 &  20.9  & 4.0 &  1  & 1 &  4.35 &  -- &  -- \\
UGC~04262  & 124.76 &  83.27  & 5.0 &  0  & 1 &  5.83 &  -- &  -- \\
NGC~2558  & 124.8 &  20.51  & 4.0 &  1  & 1 &  5.68 &  -- &  -- \\
NGC~2565  & 124.95 &  22.03  & 4.0 &  1  & 1 &  3.54 &  -- &  -- \\
NGC~2572  & 125.35 &  19.15  & 2.0 &  1  & 1 &  4.79 &  -- &  -- \\
UGC~04375  & 125.8 &  22.66  & 5.0 &  0  & 1 &  2.71 &  -- &  -- \\
NGC~2596  & 126.86 &  17.28  & 5.0 &  0  & 2 &  8.89 &  2.85 &  1.56 \\
NGC~2595  & 126.93 &  21.48  & 6.0 &  1  & 1 &  8.68 &  -- &  -- \\
NGC~2604  & 128.35 &  29.54  & 8.0 &  1  & 1 &  2.95 &  -- &  -- \\
NGC~2639  & 130.91 &  50.21  & 2.0 &  0  & 1 &  2.77 &  -- &  -- \\
NGC~2730  & 135.57 &  16.84  & 7.0 &  0  & 2 &  6.37 &  2.46 &  1.44 \\
NGC~2805  & 140.08 &  64.1  & 6.0 &  0  & 1 &  6.86 &  -- &  -- \\
NGC~2906  & 143.03 &  8.44  & 5.0 &  0  & 1 &  2.0 &  -- &  -- \\
NGC~2916  & 143.74 &  21.71  & 5.0 &  0  & 2 &  5.66 &  3.6 &  1.56 \\
UGC~05108  & 143.86 &  29.81  & 4.0 &  1  & 2 &  39.47 &  6.03 &  0.52 \\
UGC~05359  & 149.72 &  19.21  & 4.0 &  1  & 1 &  6.49 &  -- &  -- \\
UGC~05396  & 150.42 &  10.76  & 5.0 &  1  & 2 &  6.38 &  2.65 &  6.05 \\
NGC~3106  & 151.02 &  31.19  & 3.0 &  0  & 1 &  7.67 &  -- &  -- \\
NGC~3300  & 159.16 &  14.17  & 1.0 &  1  & 1 &  2.83 &  -- &  -- \\
NGC~3381  & 162.1 &  34.71  & 8.0 &  1  & 1 &  2.05 &  -- &  -- \\
IC~0674  & 167.78 &  43.63  & 3.0 &  1  & 1 &  6.29 &  -- &  -- \\
UGC~06312  & 169.5 &  7.84  & 3.0 &  0  & 1 &  6.98 &  -- &  -- \\
NGC~3614  & 169.59 &  45.75  & 5.0 &  0  & 1 &  4.83 &  -- &  -- \\
NGC~3687  & 172.0 &  29.51  & 4.0 &  1  & 1 &  2.27 &  -- &  -- \\
NGC~3811  & 175.32 &  47.69  & 5.0 &  0  & 1 &  2.59 &  -- &  -- \\
NGC~3815  & 175.41 &  24.8  & 5.0 &  1  & 1 &  2.84 &  -- &  -- \\
NGC~3994  & 179.4 &  32.28  & 5.0 &  0  & 3 &  1.27 &  3.18 &  4.3 \\
NGC~4003  & 179.5 &  23.12  & 1.0 &  1  & 1 &  6.54 &  -- &  -- \\
UGC~07012  & 180.51 &  29.85  & 7.0 &  0  & 3 &  1.75 &  2.88 &  2.84 \\
NGC~4047  & 180.71 &  48.64  & 5.0 &  0  & 3 &  2.45 &  4.13 &  3.86 \\
UGC~07145  & 182.46 &  38.22  & 5.0 &  1  & 2 &  6.33 &  2.68 &  1.68 \\
NGC~4185  & 183.34 &  28.51  & 5.0 &  1  & 2 &  7.42 &  4.66 &  1.43 \\
NGC~4210  & 183.82 &  65.99  & 4.0 &  1  & 2 &  5.69 &  2.15 &  1.32 \\
NGC~4470  & 187.41 &  7.82  & 6.0 &  0  & 1 &  1.77 &  -- &  -- \\
NGC~4711  & 192.19 &  35.33  & 5.0 &  0  & 2 &  3.95 &  2.39 &  1.85 \\
UGC~08004  & 192.91 &  31.35  & 7.0 &  0  & 2 &  6.78 &  3.02 &  2.93 \\
NGC~4961  & 196.45 &  27.73  & 7.0 &  1  & 2 &  2.56 &  1.67 &  1.79 \\
UGC~08231  & 197.16 &  54.07  & 8.0 &  1  & 1 &  2.27 &  -- &  -- \\
NGC~5000  & 197.45 &  28.91  & 5.0 &  1  & 1 &  5.58 &  -- &  -- \\
NGC~5016  & 198.03 &  24.09  & 5.0 &  0  & 1 &  2.0 &  -- &  -- \\
NGC~5157  & 201.82 &  32.03  & 3.0 &  1  & 1 &  7.32 &  -- &  -- \\
NGC~5205  & 202.51 &  62.51  & 5.0 &  1  & 1 &  1.77 &  -- &  -- \\
NGC~5267  & 205.17 &  38.79  & 3.0 &  1  & 2 &  9.76 &  2.75 &  1.08 \\
NGC~5320  & 207.58 &  41.37  & 5.0 &  1  & 1 &  3.76 &  -- &  -- \\
UGC~08781  & 208.09 &  21.54  & 4.0 &  1  & 1 &  7.66 &  -- &  -- \\
NGC~5376  & 208.82 &  59.51  & 4.0 &  0  & 2 &  2.93 &  1.68 &  1.5 \\
NGC~5379  & 208.89 &  59.74  & 3.0 &  0  & 1 &  1.82 &  -- &  -- \\
NGC~5378  & 209.21 &  37.8  & 4.0 &  1  & 1 &  5.18 &  -- &  -- \\
NGC~5406  & 210.08 &  38.92  & 4.0 &  1  & 2 &  22.01 &  5.33 &  0.51 \\
NGC~5473  & 211.18 &  54.89  & 0.0 &  1  & 3 &  2.35 &  3.61 &  1.78 \\
NGC~5480  & 211.59 &  50.73  & 7.0 &  0  & 2 &  2.95 &  1.63 &  1.14 \\
UGC~09067  & 212.69 &  15.21  & 5.0 &  0  & 1 &  4.62 &  -- &  -- \\
NGC~5520  & 213.09 &  50.35  & 5.0 &  1  & 1 &  1.65 &  -- &  -- \\
NGC~5633  & 216.87 &  46.15  & 5.0 &  0  & 1 &  1.42 &  -- &  -- \\
NGC~5657  & 217.68 &  29.18  & 5.0 &  1  & 1 &  3.96 &  -- &  -- \\
NGC~5720  & 219.64 &  50.82  & 5.0 &  1  & 1 &  6.81 &  -- &  -- \\
NGC~5732  & 220.16 &  38.64  & 5.0 &  0  & 2 &  2.58 &  1.51 &  3.75 \\
UGC~09476  & 220.38 &  44.51  & 5.0 &  0  & 2 &  3.95 &  1.51 &  2.46 \\
NGC~5784  & 223.57 &  42.56  & 0.0 &  0  & 1 &  6.65 &  -- &  -- \\
UGC~09598  & 223.79 &  43.82  & 5.0 &  0  & 2 &  5.88 &  3.04 &  2.27 \\
NGC~5876  & 227.38 &  54.51  & 1.0 &  1  & 2 &  7.15 &  2.7 &  1.66 \\
NGC~5888  & 228.28 &  41.26  & 4.0 &  1  & 2 &  9.75 &  4.55 &  3.18 \\
UGC~09777  & 228.56 &  20.48  & 5.0 &  1  & 3 &  2.49 &  4.57 &  4.53 \\
UGC~09842  & 231.27 &  37.96  & 5.0 &  1  & 1 &  6.72 &  -- &  -- \\
NGC~5957  & 233.85 &  12.05  & 4.0 &  1  & 1 &  2.53 &  -- &  -- \\
NGC~5971  & 233.9 &  56.46  & 4.0 &  0  & 2 &  4.24 &  1.38 &  2.26 \\
IC~4566  & 234.18 &  43.54  & 4.0 &  1  & 1 &  4.8 &  -- &  -- \\
NGC~5980  & 235.38 &  15.79  & 5.0 &  0  & 1 &  3.37 &  -- &  -- \\
NGC~6004  & 237.59 &  18.94  & 5.0 &  1  & 1 &  4.78 &  -- &  -- \\
IC~1151  & 239.63 &  17.44  & 7.0 &  0  & 1 &  2.59 &  -- &  -- \\
NGC~6032  & 240.75 &  20.96  & 5.0 &  1  & 2 &  29.64 &  4.69 &  0.41 \\
NGC~6060  & 241.47 &  21.48  & 4.0 &  0  & 1 &  6.65 &  -- &  -- \\
NGC~6063  & 241.8 &  7.98  & 5.0 &  0  & 2 &  4.24 &  1.87 &  1.43 \\
IC~1199  & 242.64 &  10.04  & 4.0 &  0  & 2 &  5.83 &  2.54 &  1.46 \\
NGC~6154  & 246.38 &  49.84  & 3.0 &  1  & 1 &  6.95 &  -- &  -- \\
NGC~6155  & 246.53 &  48.37  & 6.0 &  0  & 2 &  1.96 &  1.22 &  3.28 \\
NGC~6186  & 248.61 &  21.54  & 4.0 &  1  & 2 &  8.13 &  1.84 &  0.95 \\
NGC~6278  & 255.21 &  23.01  & 1.0 &  1  & 2 &  4.73 &  2.28 &  1.56 \\
NGC~6301  & 257.14 &  42.34  & 5.0 &  0  & 2 &  12.03 &  6.85 &  1.67 \\
NGC~6314  & 258.16 &  23.27  & 3.0 &  0  & 1 &  6.92 &  -- &  -- \\
UGC~10796  & 259.2 &  61.92  & 7.0 &  1  & 1 &  7.89 &  -- &  -- \\
UGC~10811  & 259.68 &  58.14  & 4.0 &  1  & 1 &  7.12 &  -- &  -- \\
IC~1256  & 260.95 &  26.49  & 4.0 &  0  & 1 &  3.59 &  -- &  -- \\
NGC~6394  & 262.59 &  59.64  & 5.0 &  1  & 2 &  7.18 &  4.32 &  2.84 \\
UGC~10905  & 263.53 &  25.34  & 1.0 &  0  & 1 &  12.36 &  -- &  -- \\
NGC~6427  & 265.91 &  25.49  & 0.0 &  1  & 1 &  2.93 &  -- &  -- \\
NGC~6478  & 267.16 &  51.15  & 6.0 &  0  & 2 &  9.58 &  5.51 &  1.31 \\
NGC~6497  & 267.82 &  59.47  & 3.0 &  1  & 2 &  8.52 &  2.87 &  1.33 \\
UGC~11228  & 276.19 &  41.49  & 0.0 &  1  & 1 &  5.21 &  -- &  -- \\
UGC~11262  & 277.65 &  42.69  & 6.0 &  0  & 2 &  6.68 &  3.67 &  1.81 \\
MCG-02-51-004  & 303.92 &  -13.62  & 4.0 &  0  & 1 &  5.19 &  -- &  -- \\
NGC~6941  & 309.1 &  -4.62  & 4.0 &  1  & 1 &  7.55 &  -- &  -- \\
NGC~6945  & 309.75 &  -4.97  & 0.0 &  1  & 1 &  4.29 &  -- &  -- \\
NGC~6978  & 313.15 &  -5.71  & 4.0 &  0  & 1 &  5.84 &  -- &  -- \\
UGC~11649  & 313.87 &  -1.23  & 3.0 &  1  & 1 &  4.01 &  -- &  -- \\
NGC~7047  & 319.12 &  -0.83  & 5.0 &  0  & 2 &  6.82 &  3.57 &  1.46 \\
NGC~7311  & 338.53 &  5.57  & 2.0 &  0  & 1 &  3.63 &  -- &  -- \\
NGC~7321  & 339.12 &  21.62  & 5.0 &  1  & 1 &  5.51 &  -- &  -- \\
UGC~12185  & 341.85 &  31.37  & 4.0 &  1  & 2 &  12.35 &  3.58 &  1.16 \\
UGC~12224  & 343.16 &  6.09  & 6.0 &  0  & 1 &  4.63 &  -- &  -- \\
NGC~7466  & 345.51 &  27.05  & 5.0 &  0  & 2 &  7.59 &  3.27 &  2.71 \\
NGC~7489  & 346.89 &  23.0  & 5.0 &  0  & 1 &  5.47 &  -- &  -- \\
NGC~7536  & 348.55 &  13.43  & 6.0 &  0  & 2 &  8.7 &  3.15 &  0.95 \\
NGC~7549  & 348.82 &  19.04  & 5.0 &  1  & 1 &  8.24 &  -- &  -- \\
NGC~7563  & 348.98 &  13.2  & 2.0 &  1  & 2 &  18.32 &  2.29 &  0.41 \\
NGC~7591  & 349.57 &  6.59  & 5.0 &  1  & 1 &  6.4 &  -- &  -- \\
IC~5309  & 349.8 &  8.11  & 6.0 &  0  & 1 &  3.68 &  -- &  -- \\
NGC~7611  & 349.9 &  8.06  & 0.0 &  1  & 1 &  2.76 &  -- &  -- \\
NGC~7623  & 350.13 &  8.4  & 0.0 &  1  & 2 &  5.21 &  3.77 &  1.24 \\
NGC~7631  & 350.36 &  8.22  & 4.0 &  0  & 2 &  3.85 &  1.97 &  3.14 \\
NGC~7653  & 351.21 &  15.28  & 4.0 &  0  & 1 &  3.76 &  -- &  -- \\
NGC~7671  & 351.83 &  12.47  & 0.0 &  1  & 3 &  2.88 &  4.71 &  4.54 \\
NGC~7691  & 353.1 &  15.85  & 5.0 &  1  & 1 &  4.95 &  -- &  -- \\
NGC~7716  & 354.13 &  0.3  & 4.0 &  1  & 1 &  4.52 &  -- &  -- \\
NGC~7722  & 354.67 &  15.95  & 3.0 &  0  & 1 &  2.91 &  -- &  -- \\
NGC~7738  & 356.01 &  0.52  & 4.0 &  1  & 1 &  4.81 &  -- &  -- \\
UGC~12810  & 357.78 &  1.06  & 5.0 &  1  & 2 &  8.35 &  2.23 &  2.4 \\
UGC~12816  & 357.96 &  3.08  & 6.0 &  0  & 2 &  11.6 &  4.41 &  1.81 \\
NGC~7782  & 358.47 &  7.97  & 4.0 &  0  & 1 &  6.65 &  -- &  -- \\
NGC~7787  & 359.03 &  0.55  & 3.0 &  0  & 1 &  4.8 &  -- &  -- \\
UGC~12864  & 359.35 &  30.99  & 6.0 &  1  & 1 &  11.85 &  -- &  -- \\
UGC~04455  & 127.89 &  -1.2  & 4.0 &  1  & 1 &  3.39 &  -- &  -- \\
UGC~06249  & 168.34 &  59.91  & 6.0 &  0  & 1 &  11.39 &  -- &  -- \\
SN2002ji  & 170.73 &  16.59  & 7.0 &  0  & 2 &  0.97 &  0.78 &  2.07 \\
UGC~07129  & 182.23 &  41.74  & 3.0 &  1  & 1 &  0.9 &  -- &  -- \\
UGC~08909  & 209.66 &  60.8  & 7.0 &  1  & 1 &  0.85 &  -- &  -- \\
NGC~0495  & 20.73 &  33.47  & 3.0 &  1  & 1 &  1.28 &  -- &  -- \\
KUG1349+143  & 207.89 &  14.11  & 5.0 &  1  & 1 &  7.23 &  -- &  -- \\
CGCG163-062  & 217.3 &  30.08  & 7.0 &  1  & 1 &  3.76 &  -- &  -- \\
NGC~5794  & 223.97 &  49.73  & 1.0 &  1  & 1 &  3.58 &  -- &  -- \\
IC~1078  & 224.12 &  9.35  & 4.0 &  1  & 2 &  4.83 &  0.7 &  2.64 \\
NGC~6977  & 313.12 &  -5.75  & 4.0 &  1  & 1 &  7.23 &  -- &  -- \\
SDSSJ015424  & 28.6 &  13.54  & 1.0 &  0  & 3 &  2.75 &  4.53 &  2.68 \\
NGC~2691  & 133.69 &  39.54  & 2.0 &  0  & 1 &  2.98 &  -- &  -- \\
NGC~2780  & 138.18 &  34.93  & 5.0 &  1  & 2 &  11.9 &  1.68 &  0.81 \\
UGC~06517  & 173.01 &  36.7  & 6.0 &  1  & 1 &  2.0 &  -- &  -- \\
NGC~5145  & 201.31 &  43.27  & 5.0 &  0  & 1 &  2.64 &  -- &  -- \\
NGC~5950  & 232.88 &  40.43  & 6.0 &  1  & 3 &  1.35 &  1.75 &  5.99 \\
UGC~10803  & 259.02 &  73.44  & 5.0 &  0  & 1 &  1.29 &  -- &  -- \\
MCG-01-52-012  & 309.46 &  -6.09  & 0.0 &  0  & 1 &  0.7 &  -- &  -- \\
UGC~09837  & 230.97 &  58.05  & 7.0 &  0  & 2 &  4.75 &  1.58 &  3.12 \\
UGC~12250  & 343.9 &  12.79  & 5.0 &  1  & 2 &  14.95 &  1.11 &  0.39 \\
NGC~5947  & 232.65 &  42.72  & 5.0 &  1  & 2 &  6.9 &  2.44 &  3.75 \\
\hline
\caption{List of galaxies. (1) Name of the galaxy; (2) right ascention (J200); (3) declination (J2000); (4) morphological type; (5) bar (1, yes; 0, no); (6) surface brightness profile type; (7) inner disc scale-length (kpc); (8) outer disc scale-length (kpc); and (9) break radius in units of h$_{\rm in}$. Columns (1), (2), (3), and (4) from the CALIFA general sample characterisation \citep[][]{2014A&A...569A...1W}. Columns (5), (6), (7), (8), (9), and (10) from the 2D decomposition (see section \ref{II_jairo}).}\\
\end{longtable}
\label{II_galaxy_tab}
\end{longtab}

\section{Photometric analysis}
\label{II_SDSS_gasp2d}

We use the fully-calibrated $g$, $r$, and $i$ band images from the SDSS seventh data release \citep[DR7, ][]{2009ApJS..182..543A} to analyse the light distribution of the sample of galaxies. We have chosen these filters to take advantage of the higher quality of the SDSS images in these bands compared to $u$ and $z$ bands. This analysis has been presented in detail in \citet[][]{2017A&A...598A..32M} covering the entire CALIFA sample, in this section we outline its main characteristics and focus on the extraction of colour profiles as a by-product of this work. We encourage the reader to check that work for further details on the DR7 SDSS photometric images as well as the sky subtraction, critical in the outermost regions of spiral galaxies.

\subsection{2D Surface-brightness distribution}
\label{II_jairo}

We perform a 2D photometric decomposition of the structural components shaping our galaxy sample by applying GASP2D \citep[][]{2008A&A...478..353M, 2014A&A...572A..25M} to the SDSS sky-subtracted images. GASP2D iteratively fits a model of the SB distribution to the pixels of the galaxy image by means of a non-linear least-squares minimization based on a robust Levenberg-Marquardt method \citep[][]{more80} using the MPFIT algorithm \citep[][]{2009ASPC..411..251M}. It weights every pixel in the image according to the variance of its photon counts, assuming a photon noise limitation and considering the detector readout noise. It deals with seeing effects by convolving the model image with a circular Moffat \citep[][]{2001MNRAS.328..977T} point spread function (PSF) with the full width at half maximum (FWHM) measured directly from stars in the galaxy image. The code allows us to determine the different photometric structures contributing to the light distribution of the studied galaxy. The components that GASP2D can fit are a bulge (or nuclear source for very small bulges), a disc (or broken disc), and/or a bar (or double bar). This way, GASP2D provides us with the set of structural parameters of these components that better fit the observed light distribution, such as ellipticities and position angles ($e$ and PA), the bar length, the break radius (R$_{\rm break}$) for galaxies with broken profiles, the inner and outer disc-scalelengths (h$_{\rm in}$, h$_{\rm out}$), etc.

\begin{figure*}
\centering
\includegraphics[width = \textwidth]{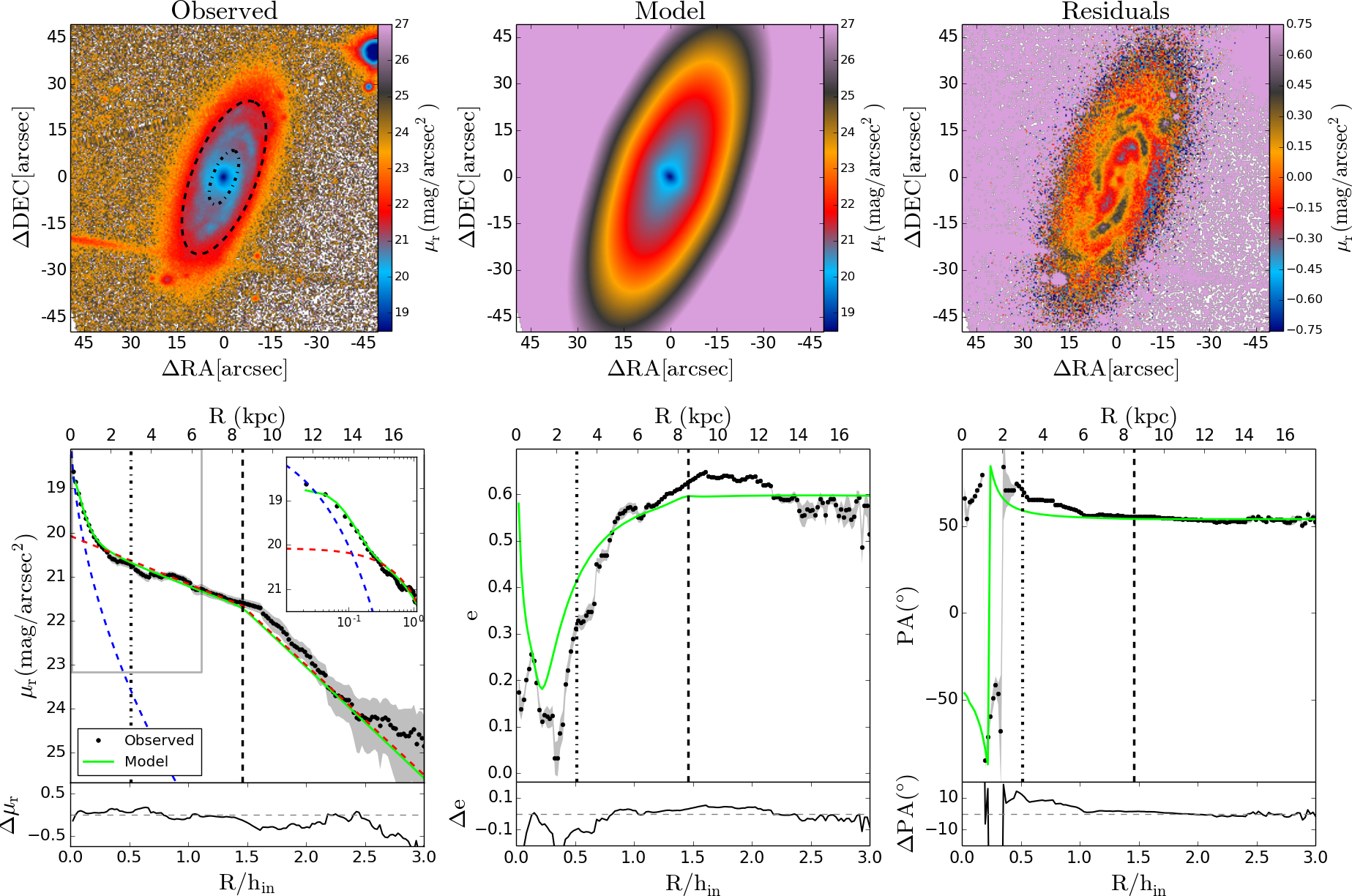} \\
\caption{Plots summarising the 2D decomposition for IC~1199. Top left: $r$-band SDSS image. Top middle: galaxy model derived from GASP2D fit considering a bulge and a broken disc component. Top right: residuals image derived by subtracting the observed image to the model. North is up in all these images. Bottom left: SDSS $r$-band surface brightness profile. Bottom middle: Ellipticity profile. Bottom right: Position angle profile. In bottom panels: points represent the observed magnitude; shadow areas account for the errors, computed as the one sigma of the distribution of values within the ellipse; red dashed line is the broken disc model; blue dashed line follows the bulge light distribution; light-green line is the output of {\tt ellipse} applied to the model; the inset in bottom left panel is focused on the inner part with a logarithmic radial scale. All bottom-panel plots show an auxiliary plot below with the residuals between observed and model values. Dotted-dashed vertical lines and the dotted-dashed ellipse delimit the bulge-dominated region. Dashed vertical lines and the dashed ellipse are located at the break radius.}
\label{II_jairo_plot}
\end{figure*}

Figure~\ref{II_jairo_plot} shows the 2D decomposition into a bulge and a broken disc components for IC~1199 as a good example of the typical performance of GASP2D in our sample of galaxies \citep[see also ][]{2017A&A...598A..32M}. Top row displays three panels with the observed SDSS $r$-band image (left), the 2D model (middle), and the residuals (model - observed, right). As an average for the entire sample, the residuals oscillate between $\pm$~0.45~mag/arcsec$^2$, with the bulk of pixels showing residuals within $\pm$~0.15~mag/arcsec$^2$ ($\sim$ 60\%). Bottom row shows the SB (left), $e$ (middle), and PA (right) profiles. In those panels black dots and shaded area represent the observed values and the errors computed as one sigma of the distribution of values within the ellipse, respectively. The solid light-green line is the output from the {\tt ellipse IRAF}\footnote{{\tt IRAF} is distributed by the National Optical Astronomy Observatory, which is operated by the Association of Universities for Research in Astronomy (AURA) under cooperative agreement with the National Science Foundation.} task applied to the model image. {\tt ellipse} fits the galaxy isophotal light distribution by means of ellipses of variable {\it e} and PA. The left panel also illustrates the contribution to the light profile coming from the bulge (dashed blue line) and the broken disc (dashed red line) components. Dotted-dashed vertical lines and the dotted-dashed ellipse in top-left panel delimit the inner region (affected by the bulge) and the beginning of the disc-dominated region (R$_{\rm lim, in}$, see Sect.~\ref{II_colours} for details). Dashed vertical lines and the dashed ellipse in the observed image are located at the break radius (R$_{\rm break}$). Despite the difficulties of fitting complex systems dominated by spiral structure or \hii~regions with smooth components, the agreement between observed and reconstructed profiles is reasonably good (see bottom row auxiliary panels).

The results from this analysis are summarised in Table~\ref{II_galaxy_tab} and have been published recently in \citet[][]{2017A&A...598A..32M}. Although it is beyond the scope of the present paper, in Appendix~\ref{II_SB_prof_class} we properly characterise the SB profiles of the galaxies analysed in this work as well as compare our results with the literature. For the purposes of the main work in this paper, we find that this sample is comprised by 132 type I, 69 type II, and 13 type III galaxies. 

\subsection{Colour profiles}
\label{II_colours}

SDSS data are also used to compute the $g-r$, $g-i$, and $r-i$ colour profiles for the sample of galaxies by running the {\tt ellipse IRAF} task to the $g$, $r$, and $i$ science frames. We fix the {\it e} and PA of the successive ellipses matching those of the  outer disc (according to the GASP2D analysis in each filter). The 1D light profiles in the three filters are calibrated in flux according to the SDSS DR7 webpage\footnote{\url{http://www.sdss2.org/dr7/algorithms/fluxcal.html}} and subtracted accordingly to obtain the three colour profiles. The errors in the colour profiles are quadratically propagated from the errors in the SB profiles of the bands involved (e.g. $g$ and $r$ bands in the case of the $g-r$ colour) computed as one sigma of the distribution of values (see Sect.~\ref{II_jairo}).

We compute linear fits to these colour profiles to quantify and describe their general trends. We restrain the fit to the disc region, avoiding the inner or bulge-dominated part. Considering that the bulge region is where the observed light distribution deviate from the disc exponential profile, we define the inner limit (R$_{\rm lim, in}$) as:
\begin{eqnarray}
 \rm \mu_{\rm disc}(r) - \mu_{\rm obs}(r) >  0.2 \,{\rm mag/arcsec}^2  \qquad  \forall  \hspace{1mm}   r < R_{\rm lim, in};
\end{eqnarray} 
where $r$ is the radius, $\mu_{\rm disc}$ is the functional shape of the disc light distribution found by GASP2D, and $\mu_{\rm obs}$ is the observed SB profile (black points in bottom-left panel of Fig.~\ref{II_jairo_plot}). The choice of 0.2 mag/arcsec$^2$ of difference between the theoretical-disc and the observed SB profiles allows us to properly avoid in this computation the region dominated by the bulge or the bar.

\begin{figure}
\centering
\includegraphics[width = 0.45\textwidth]{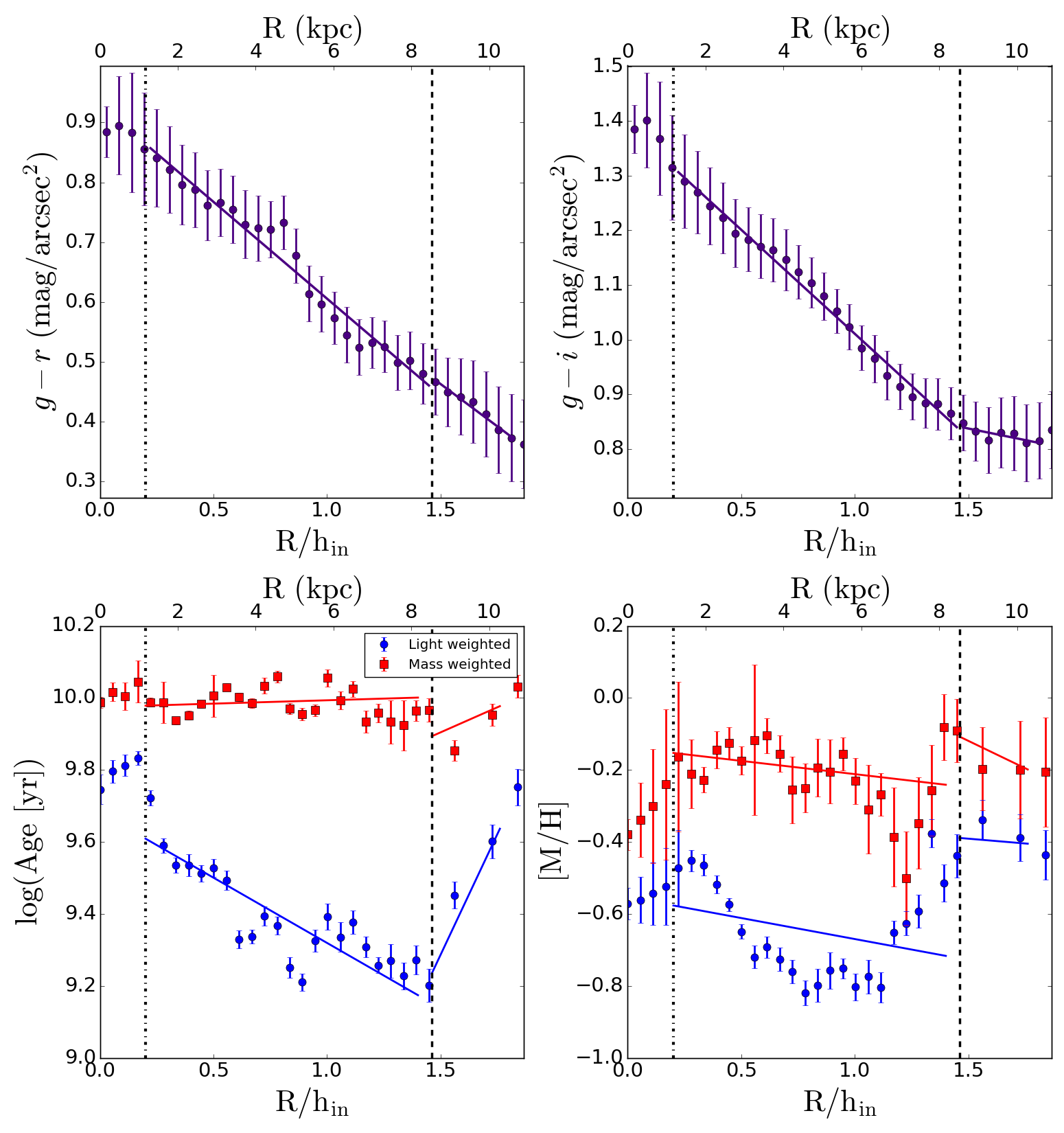} \\
\caption{$g-r$ and $g-i$ colour (top panels) and stellar age and metallicity (bottom panels) profiles for IC~1199 as an example. The colour profiles are represented by means of purple circles (top panels). Blue circles represent light-weighted stellar age or metallicity; red squares symbolise mass-weighted stellar age or metallicity (bottom panels). The dotted-dashed vertical line delimits the bulge-dominated region (R$_{\rm lim, in}$), while the dashed vertical is located at the break radius (R$_{\rm break}$).}
\label{II_colour_vs_ste_pop}
\end{figure}

For a fair comparison between colour and stellar parameters profile gradients (see Sect.~\ref{II_results}), we have decided to apply as outer limit (R$_{\rm lim, out}$) for the colour linear fits the one defined from the stellar age and metallicity profiles (see Sect.~\ref{II_age_met_grads}). We compute the inner gradients taking into account colour values from R$_{\rm lim, in}$ to R$_{\rm lim, out}$ for type I galaxies and from R$_{\rm lim, in}$ to R$_{\rm break}$ for type II and III galaxies. We highlight that we are able to compute high-quality colour profiles beyond R$_{\rm lim, out}$. However, we fix the outer limit to R$_{\rm lim, out}$ to properly compare with the stellar parameters gradients. In Fig.~\ref{II_colour_vs_ste_pop} (top panels) we show the colour profiles and the results of the linear fits for IC~1199 as an example. The performed linear fits are error-weighted and take into account the observational errors of the radial points to derive the parameters of the fit, such as the gradient and its error. The values for all the derived gradients and errors are given in Table~\ref{II_tab_colours} in Appendix~\ref{sec:appendix1}.

\section{Spectroscopic analysis}
\label{II_spec_analysis}


Our sample of galaxies was drawn from those comprising the CALIFA survey, as already stated in Sect.~\ref{II_sample}. CALIFA data (as well as the data from the CALIFA extended projects) were collected at the 3.5\,m telescope at Calar Alto using the PMAS integral field unit spectrograph in its PPaK configuration \citep{2005PASP..117..620R}. This project provides high quality spectra of 667 galaxies in the local Universe \citep[0.005 < $z$ < 0.03,][]{2014A&A...569A...1W}. Two different observing set-ups were adopted to maximise the scientific impact of the survey, one at high resolution (V1200) and the other at a lower resolution (V500). The wavelength range of the V500 (V1200) data is 3745~\AA -- 7500~\AA $ $ (3650~\AA -- 4840~\AA) with a spectral resolution of FWHM = 6.0 \AA $ $ (FWHM = 2.7 \AA). A three-position dithering scheme was chosen to obtain a 100\,\% coverage of the entire field-of-view with a final exposure time of 2700\,s for the V500 data (5400 s for the V1200 setup). We use the CALIFA COMBO cubes from the version 1.5 of the reduction pipeline \citep[][]{2015A&A...576A.135G}. The COMBO cubes are a combination of the V500 and the V1200 (degraded to the V500 resolution) datacubes that avoids the vignetting of the data in the blue end of some fibres for the V500 data \citep[see ][]{2012A&A...538A...8S}. This COMBO dataset displays high-quality spectra across the entire field-of-view in the wavelength range from 3700 \AA~to 7500 \AA.

This dataset has been proven unique to analyse the outer parts of spiral galaxies \citep[see][]{2016MNRAS.456L..35R, 2016A&A...585A..47M, 2016A&A...587A..70S} because of its wide field-of-view ($74''\times64''$). In addition, the sky subtraction, crucial for our purposes, has been thoroughly studied to maximise the information coming from the 36 PMAS sky-fibers \citep[see ][]{2013A&A...549A..87H}.

\subsection{Stellar population analysis}
\label{II_ste_pop_analysis}

The methodology applied in this work to extract the stellar age and metallicity distributions from the CALIFA data has been extensively tested in previous works \citep[e.g.][]{2011MNRAS.415..709S, 2015MNRAS.446.2837S}. In particular, in \citet[][]{2015A&A...583A..60R}, the authors apply this method to an integrated spectrum coming from the scanning of a wide region in the bar of the Large Magellanic Cloud to determine its stellar content. We compare those results to the star formation history reconstructed using state-of-the-art methods comparing synthetic and observed colour-magnitude diagrams \citep[][]{2004AJ....128.1465A, 2009AJ....138..558A, 2011ApJ...730...14H, 2010ApJ...720.1225M}. The good agreement between both approaches encourages us to apply this method to external systems, supported by other works in the literature \citep[][]{2012MNRAS.423..406G, 2016A&A...593A..78K}.

In this section we carefully explain this method. We must note here that, as a preliminary step, we mask the CALIFA datacubes to avoid foreground and background stars, as well as bad or low S/N spaxels. This methodology can be divided into three main steps:

i) Stellar kinematics: An adaptive Voronoi method following the \citet[][]{2003MNRAS.342..345C} algorithm is applied to the masked data with a goal continuum S/N of 20 and considering just spaxels with a minimum S/N of 3. We apply a stellar kinematics pipeline specifically designed for dealing with the CALIFA data \citep[][]{2017A&A...597A..48F} to these binned spectra based on the ``penalised pixel fitting'' code {\tt pPXF} \citep[][]{2004PASP..116..138C, 2011MNRAS.413..813C}. This code uses a maximum-likelihood approach to match the observed spectrum with a combination of stellar templates, once they have been convolved with a line-of-sight velocity distribution (LOSVD). The LOSVD used by {\tt pPXF} can be described via the Gauss-Hermite parametrization, allowing the measurement of the velocity, the velocity dispersion and higher order Gauss-Hermite moments up to the h3 and h4 \citep[][]{1993MNRAS.265..213G, 1993ApJ...407..525V}. To derive the velocity and velocity dispersion maps we use a subset of the INDOUSv2 library \citep[][]{2004ApJS..152..251V}. Typical errors in the velocity determination are of the order of 5 to 20 km s$^{-1}$ (for inner to outer spaxels). We use the computed velocity and velocity dispersion maps to shift the observed datacubes to the rest-frame and convolve them to a final FWHM of 8.4~\AA, resolution that is well suited for analysing low and intermediate-mass galaxies \citep[][]{2010MNRAS.404.1639V}.

Afterwards, we apply an elliptical integration (annuli) to these corrected datacubes. The width of each annulus is not fixed with the purpose of having spectra with at least a S/N of 20 (per \AA, in the continuum). The centre, ellipticity, and position angle of the ellipses are fixed, matching the outer disc isophotes from GASP2D (see section \ref{II_jairo}).

ii) Emission line removal: We use {\tt GANDALF} \citep[Gas AND Absorption Line Fitting, ][]{2006MNRAS.366.1151S, 2006MNRAS.369..529F} to decontaminate the CALIFA spectra in order to have pure absorption spectra. This code is able to simultaneously recover the stellar and ionised gas kinematics and content. {\tt GANDALF} treats emission lines as additional gaussian templates to add to the best combination of stellar templates (accounting for the stellar continuum). The code has been modified in order to take into account the dependency with wavelength of the instrumental FWHM when transformed to velocity units (L. Coccato and M. Sarzi, private communication). In addition, the wavelength range has been limited to the blue part (3800 Å to 5800 Å) as most of the spectral features sensitive to the stellar populations are located in this region of the spectrum. We use an optimal subset of the \citet{2010MNRAS.404.1639V} models (hereafter, V10) with a Kroupa universal initial mass function \citep[][]{2001MNRAS.322..231K}. These models are based on the MILES library\footnote{The models are publicly available at \url{http://miles.iac.es}} \citep{2006MNRAS.371..703S, 2011A&A...532A..95F} as observed stellar templates. The shape and position of the emission lines computed in this way is subtracted to the observed spectrum (with contribution from stars and gas) to obtain the ``pure-absorption'' stellar spectrum.

iii) Stellar content: To recover the stellar content from these ``absorption-pure'' spectra we use {\tt STECKMAP}\footnote{{\tt STECKMAP} can be downloaded at \url{http://astro.u-strasbg.fr/~ocvirk/}} \citep[STEllar Content and Kinematics via Maximum A Posteriori likelihood, ][]{2006MNRAS.365...74O, 2006MNRAS.365...46O}. {\tt STECKMAP} is aimed at simultaneously recovering the stellar content and stellar kinematics using a Bayesian method via a maximum {\it a posteriori} algorithm. It is based on the minimization of a penalised $\chi^2$ while no {\it a priori} shape of the solution is assumed (i.e. it is a non-parametric program). The definition of the penalised $\chi^2$ function to minimise is:
\begin{equation}
 \rm Q_{\mu} = {\chi^2(s(x,Z,g)) + P_{\mu}(x,Z,g)},
\label{II_Q_def}
\end{equation}
where {\it s} is the modelled spectrum which depends on the stellar content (age distribution, {\it x}; and Age-Metallicity relation, {\it Z} in the above equation) and the stellar kinematics (broadening function, {\it g}). The {\tt STECKMAP} output consists of three different solutions, the Stellar Age Distribution (SAD), the Age-Metallicity relation (AMR), and the line-of-sight velocity distribution (LOSVD). Those solutions with smooth SAD, AMR, and LOSVD are favoured while solutions with strong variations (those that are thought to be non-physical) are penalised by means of the penalisation function ($P_\mu$ in equation \ref{II_Q_def}). This penalisation function is defined as:
\begin{equation}
 \rm P_{\mu}(x,Z,g) = {\mu_{x} P(x) + \mu_{Z} P(Z) + \mu_{v} P(g)}
\label{II_P_def}
\end{equation}
The different smoothing parameters ($\mu_{\rm x}$, $\mu_Z$, and $\mu_{\rm v}$) allow the user to choose the smoothness for the different solutions (SAD, AMR, and LOSVD). The smoothness is completely accomplished by means of the function P. This function gives higher penalisation values to strongly oscillated functions while low values to smooth solutions. Higher values for the smoothing parameters as input parameters implies that more smoothed solutions are preferred. The function P can also adopt different shapes \citep[for further information see][]{2006MNRAS.365...46O, 2006MNRAS.365...74O}. {\tt STECKMAP} uses a polynomial to deal with the shape of the continuum, thus, avoiding flux calibration and extinction errors.

Prior to running {\tt STECKMAP} to the CALIFA data we investigate which combination of input parameters better fits the CALIFA data. According to those tests, we decide to use a square laplacian smoothing kernel for the shape of the function P for the SAD and AMR solutions, with values $\mu_x$ $=$ 0.01 and $\mu_Z$ $=$ 100, respectively. For further information about these parameters and the choice of them we encourage the reader to check previous works \citep[]{2006MNRAS.365...74O, 2006MNRAS.365...46O, 2010ApJ...709...88O, 2014A&A...570A...6S}. As we study the LOSVD with especially devoted codes (see above), we do not allow {\tt STECKMAP} to fit the kinematics. We convolve all the observed spectra to a common velocity dispersion of 8.4~\AA~and shift them to the rest-frame and we fix the kinematics while running {\tt STECKMAP} to these values to reduce the velocity dispersion-metallicity degeneracy reported in \citet[][]{2011MNRAS.415..709S}. In this step we use the entire set of the V10 models with ages ranging from 63 Myr to 17.8 Gyr and metallicities from -2.32 to +0.2 ([M/H]) as well as the same wavelength range that for the emission line removal (3800~\AA~to 5800~\AA).

\begin{figure}
\centering
\includegraphics[width = 0.45\textwidth]{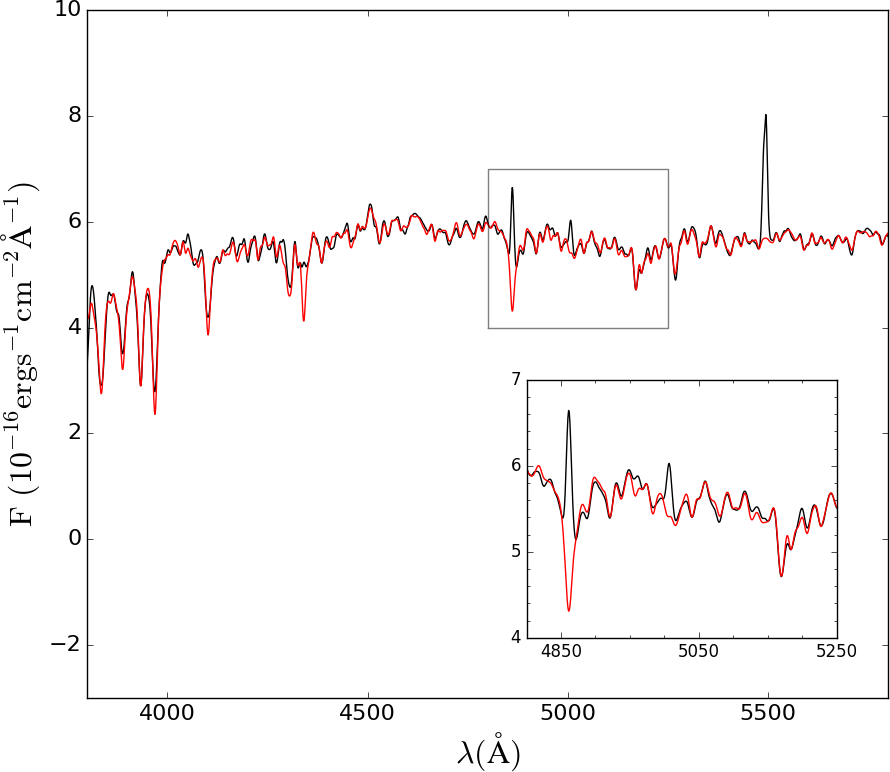} \\
\caption{Example of a typical CALIFA spectrum analysed with the described method. In particular, this is the spectrum of IC~1199 at 24 arcsecs (1.34 h$_{\rm in}$). The solid black line represents the observed, rest-framed spectrum. The solid red line is the {\tt STECKMAP} best fit to the observed spectrum after removing the gaseous emission lines with {\tt GANDALF}. The inset is focused on the H$\beta$-MgI (4800~\AA -- 5250~\AA). See text for further details.}
\label{II_spec_example}
\end{figure}

Figure~\ref{II_spec_example} shows an example of a typical, high-quality spectrum analysed with this methodology. The kinematics correction applied to the CALIFA datacubes and the elliptical integration allow us to obtain high quality spectra even in the outer parts of the analysed galaxies. In this particular case, we are showing a spectrum (solid black line) at an intermediate-to-large galactocentric distance from IC~1199 (24 arcsecs, 1.34 h$_{\rm in}$). We must note the high quality of the spectrum (S/N = 41.3 per pixel in the continuum), highlighting specially some features such as the D4000 break, emission lines (H$\beta$, \oiii, etc), or absorption features (\caii, MgI, etc). We can observe some remainings of the CALIFA sky subtraction in the \oi~sky line at 5577~\AA, region that is masked and thus, not considered in the {\tt STECKMAP} fit. The solid red line shows the {\tt STECKMAP} fit to the emission-free spectrum (from {\tt GANDALF}). The inset is focused on the region where H$\beta$ and MgI absorption features are located. 

\subsubsection{Age and metallicity radial profiles}
\label{II_age_met_grads}

To compute the age and metallicity profiles, both light-weighted (L-W) and mass-weighted (M-W), we average the AMR and the SAD from {\tt STECKMAP} in logarithmic scale, as extensively done in the literature \citep[e.g.][]{2011MNRAS.415..709S, 2013A&A...557A..86C, 2014A&A...570A...6S, 2015A&A...581A.103G}:
\begin{equation}
 \rm <{\log}({\rm Age} [{\rm yr}])>_{{\rm M-W}} = {\sum_i {\rm mass}(i)*{\log}({\rm Age}_i) \over \sum_i {\rm mass}(i)}
\end{equation}
\begin{equation}
 \rm <{\log}({\rm Age} [{\rm yr}])>_{{\rm L-W}} = {\sum_i {\rm flux}(i)*{\log}({\rm Age}_i) \over \sum_i {\rm flux}(i)}
\end{equation}
\begin{equation}
 \rm <{\rm [M/H]}>_{{\rm M-W}} = {\sum_i {\rm mass}(i)*{\log}({\rm Z}_{i}/{\rm Z}_\odot) \over \sum_i {\rm mass}(i)}
\end{equation}
\begin{equation}
 \rm <{\rm [M/H]}>_{{\rm L-W}} = {\sum_i {\rm flux}(i)*{\log}({\rm Z}_{i}/{\rm Z}_\odot) \over \sum_i {\rm flux}(i)}
\end{equation}
where Z$_\odot$ is the solar metallicity (0.02) and mass(i) and flux(i) are the mass and flux of the population with age = Age$_i$ and metallicity = Z$_{i}$. Errors in all of the above defined quantities are computed by means of 25 Monte Carlo simulations, number that has been proven sufficient to provide reasonable errors \citep[][]{2015MNRAS.446.2837S, 2016MNRAS.456L..35R}. We derive these errors by adding some noise consistent with the quality of the original data to the best fit of the observed spectrum and running {\tt STECKMAP} 25 times to all of these noisy, best fit spectra. The standard deviation of the 25 recovered age and metallicity values is what we consider the error in each magnitude.

Once we have derived the age and metallicity radial distributions, we compute linear fits in a similar way as we explained for the colour profiles (see Sect.~\ref{II_colours}). We use the same definition of R$_{\rm lim, in}$ as before and the definition of R$_{\rm lim, out}$ is imposed by the quality of the CALIFA spectroscopic data or the method performance. Thus, R$_{\rm lim, out}$ is given by the last radial spectrum with a S/N > 20 or by the last radial spectrum from which reliable stellar population results are drawn. We highlight that in a small number of cases (corresponding to the outermost regions), although the computed S/N is higher than 20 a visual inspection of the {\tt GANDALF} or {\tt STECKMAP} fits suggests that they are not good enough (bad emission line removal, deficient continuum shape reconstruction, etc) and thus, non-reliable stellar population results are obtained. The information from these spectra is discarded. For type I galaxies we compute a single (inner) gradient from R$_{\rm lim, in}$ to R$_{\rm lim, out}$ while for type II and III galaxies the outer limit is restricted to the break radius. In this case, again, the performed linear fits are error-weighted, taking into account the observational errors of the points to derive the gradient and its error. Figure~\ref{II_colour_vs_ste_pop} (bottom panels) shows an example of the typical age and metallicity profiles that we obtain with the CALIFA data. The values of all the gradients and their errors for the age and metallicity profiles (light- and mass-weighted) are given in Tables~\ref{II_tab_ages} and~\ref{II_tab_mets} in Appendix~\ref{sec:appendix1}.

\section{Results}
\label{II_results}

The procedure explained in the previous sections allows us to characterise the light distribution for the 214 galaxies under analysis and to obtain their colour profiles as well as their stellar age and metallicity profiles. In this paper we focus on the behaviour of such tends in the inner parts to identify differences in their inner gradients for galaxies displaying different SB profiles. The stellar content for some of the galaxies in this sample in the outer parts (i.e. beyond the break radius for type II galaxies or beyond three disc-scalelengths for type I galaxies) has been presented in \citet[][]{2016MNRAS.456L..35R}.

We know from previous works that stellar radial migration exists in spiral galaxies \citep[][]{2001A&A...377..911F, 2004A&A...418..989N, 2008ApJ...684L..79R, 2009MNRAS.398..591S, 2010ApJ...722..112M, 2014A&A...565A..89B, 2015MNRAS.447.3526K}. Moreover, if radial redistribution of material is important, it must have an effect in the stellar content of the inner and the outer regions. As a consequence, if there is a relation between radial migration and SB profiles, galaxies displaying different SB profiles should display differences in their inner stellar content. Therefore, the analysis of the inner colour profiles (Sect.~\ref{II_colour_gradients}) along with the inner stellar age and metallicity gradients (Sect.~\ref{II_ste_pops_gradients}) segregating in the different SB profile types can help us to shed light into the role of radial migration in the shaping of stellar parameter and light distribution profiles.

\subsection{Inner colour gradients}
\label{II_colour_gradients}

The analysis of the light distribution carried out along this work (see Sect.~\ref{II_SDSS_gasp2d}) allows us to obtain colour radial profiles. Although we know that drawing stellar population information from optical colours might not be an optimal approach \citep[][]{2009MNRAS.395.1669G}, we have studied the colour inner gradient distribution for type I, II, and III galaxies, for the sake of completeness.

\begin{table*}
\centering
\begin{tabular}{lccc}
\hline
\noalign{\vspace{0.1 truecm}}
 & \multicolumn{3}{c}{SB profile}\\ 
 & Type II & Type I & Type III \\ 
 \noalign{\vspace{0.1 truecm}}
 \hline \hline 
 \noalign{\vspace{0.1 truecm}}
$g-r$ &    $-0.12$~$\pm$~0.02 (0.16) &     $-0.084$~$\pm$~0.007 (0.08) &  $-0.04$~$\pm$~0.02 (0.07) \\
$g-i$ &    $-0.17$~$\pm$~0.03 (0.20) &     $-0.123$~$\pm$~0.010 (0.12) &  $-0.05$~$\pm$~0.02 (0.08) \\
$r-i$ &    $-0.05$~$\pm$~0.01 (0.13) &     $-0.041$~$\pm$~0.005 (0.07) &  $-0.01$~$\pm$~0.01 (0.04) \\
\noalign{\vspace{0.1 truecm}}
\hline
\end{tabular}
\caption[Colour inner gradients (averaged values)]{Error-weighted average, its error and dispersion (the later in parenthesis) values of the inner gradients for the $g-r$, $g-i$, and $r-i$ colour profiles (see text for details). Units are in mag/h$_{\rm in}$.}
\label{II_colour_av_gradients}
\end{table*}

\begin{figure*}
\centering
\includegraphics[width = 0.93\textwidth]{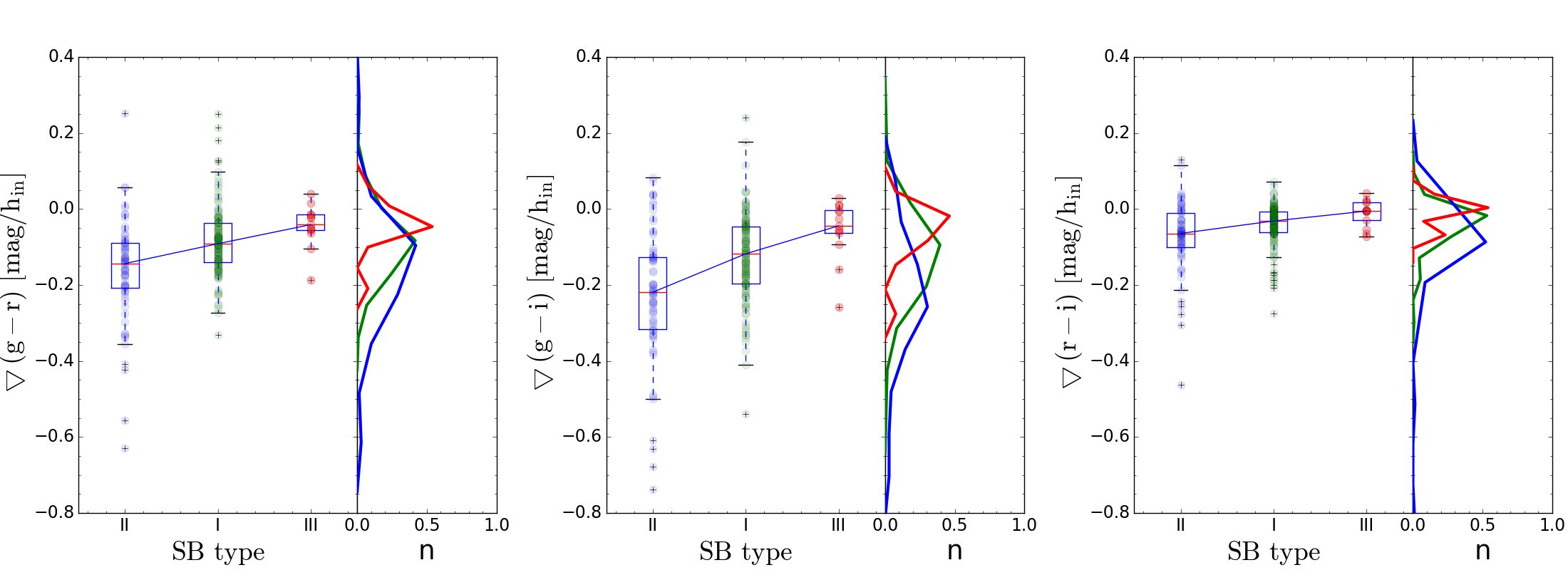} \\
\caption{Distribution of the inner $g-r$ (left), $g-i$ (middle) and $r-i$ (right) colour gradients as a function of the surface brightness profiles of the galaxies represented as box plots. The boxes extend from the lower to upper quartile values of the gradient distributions, with a red line at the median. The whiskers extend from the box to show the range of the data. Outlier points are those past the end of the whiskers and are represented with crosses. We also show the histograms for each distribution colour coded according to the SB type: type I (green), type II (blue), and type III (red). Dots with transparency represent all the observational values. The median values of all the distributions are linked using a blue line to highlight the tendency.}
\label{II_colour_inner_fig}
\end{figure*}

Figure~\ref{II_colour_inner_fig} shows the distribution of the inner gradients for the $(g-r)$, $(g-i)$, and $(r-i)$ colour profiles as box plots (main plots) as well as by means of histograms (right-hand auxiliary panels). The error-weighted mean values of the inner gradient distribution for the $(g-r)$, $(g-i)$ and $(r-i)$ colour profiles are shown in Table~\ref{II_colour_av_gradients}. We can see that gradient distributions for type I, II and III galaxies are different. Type II galaxies present the steepest gradients of all, followed by type I and type III systems (displaying the shallowest profiles). To further quantify such a result, we make use of two different statistical tests to check whether these distributions are drawn or not from the same one: the Kolmogorov-Smirnov (KS) and the Anderson-Darling (AD) tests. These tests are used to compare two samples and provide estimations on how different these two distributions are by means of a parameter called ``p-value''. If this statistical parameter is below a significance level (generally $\sim$~0.05) then, we can reject the ``null hypothesis'', i.e. we can conclude that the two samples are drawn from different distributions. The corresponding ``p-values'' for all these tests can be found in Table~\ref{ps_colours}. Very low p-values are found using the KS tests when comparing the colour inner gradient distributions for type I and II galaxies as well as comparing type I and type III distributions, and type II and III. Although the size of the sample under analysis in this work is large enough as to be analysed using a KS test, suited for large samples, we have decided to use the AD test too, which is better suited for small samples. The p-values obtained using this second statistical test (AD) agree with the ones found using KS, with p-values always well below the significance level. Then, we can conclude that the colour profile inner gradient distributions for type I, II, and III galaxies seem to be drawn from different distributions with type II galaxies displaying steeper negative trends than type I and III systems, with the latter showing the shallowest profiles.

\begin{table*}
\centering
\small
\begin{tabular}{lccc}
\hline
 SB profiles & $(g-r)$ & $(g-i)$ & $(r-i)$ \\ \hline \hline
I vs. II & 0.001 (0.0001) &  0.0001  (0.00002) &   0.0001 (0.0005)  \\
I vs. III & 0.00001 (0.012) &  0.0   (0.01) &  0.0   (0.04) \\
II vs. III & 0.0  (0.0003) &   0.0   (0.0004) & 0.0 (0.007) \\
\hline
\end{tabular}
\caption{Statistical ``p-values'' for the KS (AD tests in parenthesis) when comparing the inner gradients distributions for type I vs. type II, type I vs. type III and type II vs. type III galaxies of the colour profiles analysed in this work ($g-r$, $g-i$, and $r-i$).}
\label{ps_colours}
\end{table*}

\subsection{Inner stellar parameters gradients}
\label{II_ste_pops_gradients}

\begin{figure*}
\centering
\includegraphics[width = 0.93\textwidth]{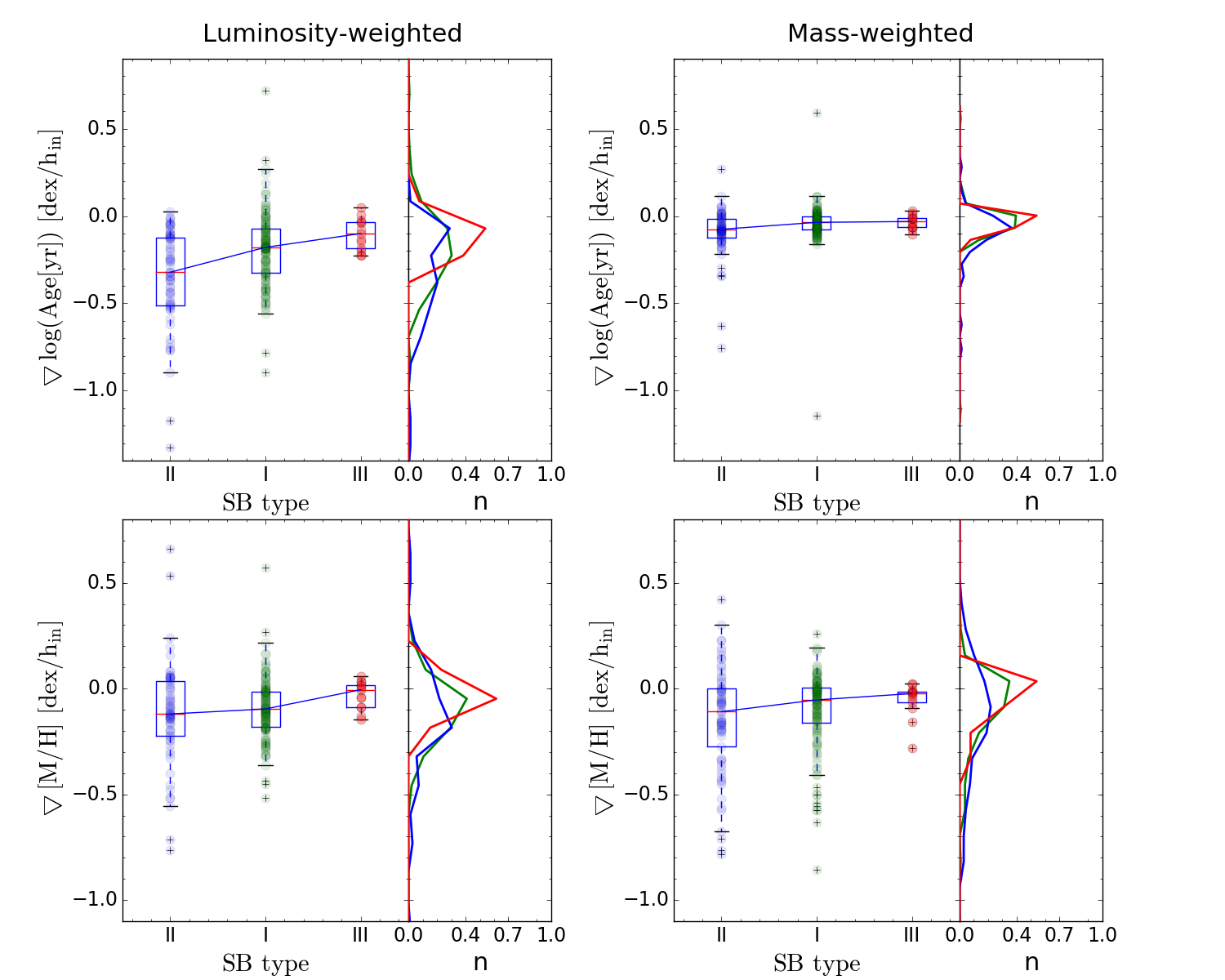} \\
\caption{Distribution of the inner gradients of the log(age[yr]) and [M/H] profiles as a function of the surface brightness profiles of the galaxies represented as box plots. The boxes extend from the lower to upper quartile values of the gradient distributions, with a red line at the median. The whiskers extend from the box to show the range of the data. Outlier points are those past the end of the whiskers and are represented with crosses. We also show the histograms for each distribution colour coded according to the SB type: type I (green), type II (blue), and type III (red). Top-left: Luminosity-weighted age; Top-right: Mass-weighted age; Bottom-left: Luminosity-weighted [M/H]; Bottom-right: Mass-weighted [M/H]. Dots with transparency represent all the observational values. The median values of all the distributions are linked using a blue line to highlight the tendency.}
\label{II_ste_inner}
\end{figure*}

To further test our hypothesis of radial migration affecting differently the inner stellar content of galaxies with different SB profiles, Fig.~\ref{II_ste_inner} shows the distribution of the inner gradients of the log(age[yr]) and [M/H] profiles (L-W and M-W) for galaxies with different SB profiles in a similar way as presented in Fig.~\ref{II_colour_inner_fig} for the colour gradients. After a visual inspection, we can claim that similar trends to the ones shown by the colour profiles are found (especially in the case of L-W quantities), though in this case those trends are not as clear as in the previous case. The observed tendency suggests that as we move from type II galaxies to type I and type III systems the slope of the L-W profiles becomes shallower. The trend is also observed but to a lesser degree in the case of the M-W quantities. The actual values for the error-weighted average gradients and their dispersions for each SB profile type are shown in Table~\ref{II_stellar_av_gradients}.

\begin{table*}
\centering
\small
\begin{tabular}{lccc}
\hline
 & \multicolumn{3}{c}{SB profile}\\
 & Type II & Type I & Type III \\ \hline \hline
L-W age &           -0.30~$\pm$~0.06 (0.51) &   -0.18~$\pm$~0.02 (0.26) &  -0.09~$\pm$~0.03  (0.11) \\
L-W metallicity &   -0.15~$\pm$~0.06 (0.47) &   -0.09~$\pm$~0.02 (0.18) &   -0.04~$\pm$~0.02 (0.06) \\
M-W age &           -0.06~$\pm$~0.02 (0.14) &   -0.03~$\pm$~0.01 (0.13) &  -0.03~$\pm$~0.02  (0.06) \\
M-W metallicity &   -0.14~$\pm$~0.03 (0.26) &   -0.09~$\pm$~0.02 (0.26) &  -0.06~$\pm$~0.02  (0.08) \\
\hline
\end{tabular}
\caption[Stellar population inner gradients (averaged values)]{Error-weighted average, its error and dispersion (the later in parenthesis) values of the inner gradients for the stellar age and metallicity profiles (see text for details). L-W stands for light-weighted quantities whereas M-W stands for mass-weighted. Units are in dex/h$_{\rm in}$.}
\label{II_stellar_av_gradients}
\end{table*}

The average values confirm the first visual impression: type II galaxies present the steepest profiles and type III the shallowest, with type I spirals displaying an intermediate behaviour in L-W parameters. However, although a similar trend is found for the gradients from the M-W quantities, we find the exception of the M-W age gradients, where type I and III galaxies display similar average values. As done in the case of the colour gradients, and to further quantify such a result, we make use of the KS and AD statistical tests. The ``p-values'' for all these tests can be found in Table~\ref{ps_pops}. KS tests seem to indicate that the observed distributions for type I and III galaxies are basically identical in the case of the M-W quantities, with high p-values (0.62 and 0.37, respectively). However, slight differences are found in the case of the L-W age gradients (p-value of 0.07, around the significance level) and clear differences in the case of the L-W metallicity gradients (p-value of 0.04, below the significance level). The AD tests arise similar results, i.e. the samples for type I and III galaxies follow the same distribution in the case of M-W quantities but differences seem to appear when comparing the distributions of the L-W gradients. Regarding the distributions shown by type II galaxies, KS and AD tests suggest that they are drawn from different distributions than type I and III systems if we pay attention to both, L-W and M-W profiles. However, this statement should be taken with caution as in the case of the metallicity gradients the p-values are in some cases similar to or even slightly above the significance level and thus, the observed differences, if any, are almost negligible.

We also reproduced the analysis shown in Fig.~\ref{II_ste_inner} but distinguishing between barred and unbarred systems instead of SB profile types to assess the effect of the presence of a bar in shaping the stellar parameter profiles. The distributions found for barred and unbarred galaxies are completely consistent among them, thus suggesting that presence of bars do not seem to have a strong effect in shaping the stellar parameter profiles \citep[in agreement with the results found in][]{2014A&A...570A...6S, 2016A&A...585A..47M}.

\begin{table*}
\centering
\small
\begin{tabular}{lcccc}
\hline
 SB profiles & L-W age & M-W age & L-W [M/H] & M-W [M/H] \\ \hline \hline
I vs. II & 0.005 (0.0003) &  0.001  (0.004) &  0.07  (0.05)  &  0.07   (0.06) \\
I vs. III & 0.07 (0.07) &  0.62   (0.60) &   0.04  (0.02) &  0.37   (0.23) \\
II vs. III &  0.001 (0.001) &    0.02  (0.03) &   0.03  (0.03) &   0.08  (0.05) \\
\hline
\end{tabular}
\caption{Statistical ``p-values'' for the KS (AD tests in parenthesis) when comparing the inner gradients distributions for type I vs. type II, type I vs. type III and type II vs. type III galaxies of the stellar properties analysed in this work (L-W and M-W age as well as L-W and M-W metallicity).}
\label{ps_pops}
\end{table*}

The similar behaviours of the inner gradients of the stellar parameter and colour profiles for type I, II, and III galaxies, especially in the case of the L-W quantities, can be interpreted as a reinforcement of the results above outlined. In addition, it might shed some light into whether optical colours can be used as a proxy for stellar ages or metallicities when looking for statistical trends in large samples of galaxies. However, in order to do such a claim further investigation and comparison between colour and stellar parameter profiles are needed.

It is worth mentioning the fact that, apart from the above mentioned differences, the distributions of inner gradients (colours and stellar parameters) also present different dispersions (see parenthesis in Tables~\ref{II_colour_av_gradients} and~\ref{II_stellar_av_gradients}). The values of the inner gradients for type III galaxies always present lower dispersions than type I or II systems. However, this is understandable considering the low number of galaxies exhibiting a type III profile (13). More intriguing is the fact that, even though the number of type I galaxies is much larger than the number of type II galaxies (132 vs. 69), type II galaxies present larger dispersion for all analysed quantities. Whether this is a consequence of the different possible processes shaping the type II SB profiles or a direct consequence of the lower degree of mixing induces by radial migration needs to be further investigated.

\section{Discussion}
\label{II_discussion}

Our hypothesis states that, i) if migration is capable of redistributing a sufficient mass fraction as to affect the observed inner stellar content and ii) different radial migration efficiencies affect the shaping of the outer light distributions, galaxies with different SB profiles should display differences in their inner colour and stellar age and metallicity gradients. The results presented in this paper suggest that this can be the case: Type II galaxies might present the lowest degree of radial migration causing the stellar population gradients to be steeper, while type III systems might display a larger degree of mixing, resulting in shallower gradients. The fact that type I systems tend to display average values in between type II and III systems and slightly different distributions to both types further supports this scenario. In this work, we have also checked if the differences in the stellar parameter gradients for type II galaxies (steep gradients) are due to a more efficient current star formation in these systems. The analysis of the light fraction of stars younger than 150 Myr yields no differences between type I, II, and III galaxies discarding a recent star formation burst as the cause of the steeper L-W age profiles in type II systems. Although tempted to claim that these results are evidences of different radial migration efficiencies for galaxies displaying different SB profiles, we must bear in mind that other processes involved in the formation of spiral galaxies besides migration can be also at play \citep[as suggested in][]{2016MNRAS.456L..35R}.  

Other theoretical and observational works have also tried to understand the role of radial migration on shaping light distribution and the stellar content in spiral galaxies. \citet[][]{2008ApJ...675L..65R} used their SPH+N-body simulations to propose that SB and age profiles were intimately related to radial migration, i.e. radial migration, along with a radial cut-off of the star formation, was the cause of the downbending of the light profile and the ageing of the outer disc. In \citet[][]{2016A&A...586A.112R}, the authors analyse the origin of the outer ageing in a set of simulated spiral galaxies \citep[RaDES,][]{2012A&A...547A..63F} finding that type I, II, and III galaxies show ``U-shape'' age profiles even when radial migration is not considered. Consistently with that work, \citet[][]{2016MNRAS.456L..35R} analysed spectroscopic data from the CALIFA survey to obtain ``U-shape'' age profiles indistinctly for type I and II galaxies suggesting that the shaping of the SB and the outer ageing on the stellar age profiles are unconnected. In addition, recent observational works suggest the existence of old outer discs formed through radial migration and/or satellite accretion in a wide variety of galactic systems. \citet[][]{2012AJ....143...47Z} analysing broad-band images of 34 dwarf irregular galaxies found some hints of old stellar populations over the entire disc from their spectral energy distribution (SED) modelling. Similar results have been found by \citet[][]{2015ApJ...800..120Z} analysing SED modelling of 698 spiral galaxies from the Pan-STARRS1 Medium Deep Survey images. \citet[][]{2016AJ....151....4D}, analysing deep optical and near-infrared images of a sample of fifteen nearby spiral galaxies, found results consistent with an inside-out disc formation coupled with an old stellar outer disc. The analysis of deep colour magnitude diagrams also claim the presence of old stars in the outskirts of very nearby systems \citep[][]{2012MNRAS.420.2625B, 2012ApJ...753..138R, 2015MNRAS.446.2789B}. 

The variety of results found in the previous mentioned works suggest that we are still far from understanding the real role of radial migration in shaping the stellar distribution (light, age, and metallicity) of spiral galaxies. Although it seems that radial migration is present in galaxies, we still cannot decouple the possible effects of recent star formation, satellite accretion or radial migration in modelling observed galaxies. The observed properties can be the consequence of a complex process, involving not only radial migration but multicomponent star formation recipes \citep[][]{2006ApJ...636..712E}, combinations of star formation thresholds and collapse of the protogalactic cloud conserving angular momentum \citep[][]{2001MNRAS.327.1334V}, satellite accretion, etc.

However, the results shown in this paper are in agreement with the speculations presented in \citet[][]{2009MNRAS.398..591S} and might indicate that there can be a radial migration efficiency transition between type II, I, and III. If this is the case, the factor determining the type of SB profile to exhibit would be the amount of outward migrated stars. If outward radial migration is high enough, this migration could change the observed SB profiles of spiral galaxies from type II to I or even III (with the combined action of satellite accretion).

This work helps us to constraint future theoretical works studying the origin of the outskirts of spiral galaxies and propose a possible scenario to explain the different observed SB profiles. Although we cannot conclude whether radial migration is the main mechanism or it is just one of many mechanisms shaping the different SB profiles, we can claim that the amount of stars that are currently displaced from their birth locations can represent a significant fraction of the total stellar mass as to affect the inner stellar age and metallicity profiles.

\section{Conclusions}
\label{II_conclusions}

In this work we present stellar age and metallicity profiles from full-spectrum fitting analysis of the CALIFA IFS data of a sample of 214 spiral galaxies. We have carefully analysed their 2D light distribution to characterise their SB profiles and obtained colour profiles from the analysis of SDSS data. Our main conclusions state that:

i) There are statistical differences in the behaviour of the colour profiles ($g-r$, $g-i$ and $r-i$) in the inner parts for type II galaxies compared to type I and III systems. Type II galaxies tend to display the steepest profiles and type III the shallowest while type I galaxies show an intermediate behaviour.

ii) Similar findings are obtained when computing the inner gradients of the L-W stellar age and metallicity trends in the inner parts. The trend is also reproduced in the case of M-W quantities but to a lesser extent.

iii) Conclusions i) and ii) together suggest a segregation in the radial migration efficiency for type I, II, and III galaxies, with type III systems being the most efficient (shallowest profiles) and type II being the less efficient systems (steepest profiles). These findings allow us to suggest that radial migration seems to have an effect in shaping the stellar content in spiral galaxies (light, stellar age and stellar metallicity distributions). The efficiency of the outward radial migration might be the factor determining the observed SB profile in a galaxy, from type II (low efficiency) to type III (high efficiency).  However, a joint effort of simulations and observations is needed in order to properly asses: a) the actual role of radial migration in galaxy evolution and its causes, b) the role of satellite accretion, and c) whether these different behaviours between type I, II and III systems are imprinted at birth or really acquired via radial migration. 

\begin{acknowledgements}
We thank the referee for very useful suggestions and comments that have helped improve the current version of this manuscript. The authors also thank L. Coccato and M. Sarzi for the new version of the {\tt GANDALF} code taking into account the dependency with wavelength of the instrumental FWHM. We would like to thank Reynier Peletier and St\'ephane Courteau for very helpful suggestions and discussions that have considerably improved this paper. This paper is (partially) based on data obtained by the CALIFA Survey, funded by the Spanish Ministery of Science under grant ICTS-2009-10, and the Centro Astron\'omico Hispano-Alem\'an. TRL thanks the support of the Spanish Ministerio de Educaci\'on, Cultura y Deporte by means of the FPU fellowship. This research has been supported by the Spanish Ministry of Science and Innovation (MICINN) under grants AYA2014-53506-P, AYA2007-67625-C02-02, AYA2014-56795-P and Consolider-Ingenio CSD2010-00064; and by the Junta de Andaluc\'ia (FQM-108). L.G. was supported in part by the US National Science Foundation under Grant AST-1311862. IM would like to thank support under grants  AYA2013-42227-P and AYA2016-76682-C3-1-P. S.Z. has been supported by the EU Marie Curie Career Integration Grant SteMaGE Nr. PCIG12-GA-2012-326466 (Call Identifier: FP7-PEOPLE-2012 CIG). JFB thanks the support received under grant AYA2016-77237-C3-1-P. AdLC acknowledges support from the CONACyT-125180, DGAPA-IA100815 and DGAPA-IA101217 projects. RAM acknowledges support by the Swiss National Science Foundation. JMA thanks support from the support from the European Research Council Starting Grant (SEDmorph;  P.I. V. Wild) and MINECO through the grant AYA2013-43188-P. 

Funding for the creation and distribution of the SDSS Archive has been provided by the Alfred P. Sloan Foundation, the Participating Institutions, the National Aeronautics and Space Administration, the National Science Foundation, the US Department of Energy, the Japanese Monbukagakusho, and the Max Planck Society. The SDSS Web site is \url{http://www.sdss.org/}. The SDSS is managed by the Astrophysical Research Consortium (ARC) for the Participating Institutions. The Participating Institutions are The University of Chicago, Fermilab, the Institute for Advanced Study, the Japan Participation Group, The Johns Hopkins University, the Korean Scientist Group, Los Alamos National Laboratory, the Max-Planck-Institute for Astronomy (MPIA), the Max-Planck-Institute for Astrophysics (MPA), New Mexico State University, University of Pittsburgh, University of Portsmouth, Princeton University, the United States Naval Observatory, and the University of Washington.
\end{acknowledgements}


\bibliography{bibliography} 

\begin{thebibliography}{124}
\expandafter\ifx\csname natexlab\endcsname\relax\def\natexlab#1{#1}\fi

\bibitem[{{Abazajian} {et~al.}(2009){Abazajian}, {Adelman-McCarthy},
  {Ag{\"u}eros}, {Allam}, {Allende Prieto}, {An}, {Anderson}, {Anderson},
  {Annis}, {Bahcall}, \& et~al.}]{2009ApJS..182..543A}
{Abazajian}, K.~N., {Adelman-McCarthy}, J.~K., {Ag{\"u}eros}, M.~A., {et~al.}
  2009, \apjs, 182, 543

\bibitem[{{Aparicio} \& {Gallart}(2004)}]{2004AJ....128.1465A}
{Aparicio}, A. \& {Gallart}, C. 2004, \aj, 128, 1465

\bibitem[{{Aparicio} \& {Hidalgo}(2009)}]{2009AJ....138..558A}
{Aparicio}, A. \& {Hidalgo}, S.~L. 2009, \aj, 138, 558

\bibitem[{{Beasley} {et~al.}(2015){Beasley}, {San Roman}, {Gallart},
  {Sarajedini}, \& {Aparicio}}]{2015MNRAS.451.3400B}
{Beasley}, M.~A., {San Roman}, I., {Gallart}, C., {Sarajedini}, A., \&
  {Aparicio}, A. 2015, \mnras, 451, 3400

\bibitem[{{Bell} \& {de Jong}(2000)}]{2000MNRAS.312..497B}
{Bell}, E.~F. \& {de Jong}, R.~S. 2000, \mnras, 312, 497

\bibitem[{{Bergemann} {et~al.}(2014){Bergemann}, {Ruchti}, {Serenelli},
  {Feltzing}, {Alves-Brito}, {Asplund}, {Bensby}, {Gruyters}, {Heiter},
  {Hourihane}, {Korn}, {Lind}, {Marino}, {Jofre}, {Nordlander}, {Ryde},
  {Worley}, {Gilmore}, {Randich}, {Ferguson}, {Jeffries}, {Micela},
  {Negueruela}, {Prusti}, {Rix}, {Vallenari}, {Alfaro}, {Allende Prieto},
  {Bragaglia}, {Koposov}, {Lanzafame}, {Pancino}, {Recio-Blanco}, {Smiljanic},
  {Walton}, {Costado}, {Franciosini}, {Hill}, {Lardo}, {de Laverny}, {Magrini},
  {Maiorca}, {Masseron}, {Morbidelli}, {Sacco}, {Kordopatis}, \& {Tautvai{\v
  s}ien{\.e}}}]{2014A&A...565A..89B}
{Bergemann}, M., {Ruchti}, G.~R., {Serenelli}, A., {et~al.} 2014, \aap, 565,
  A89

\bibitem[{{Bernard} {et~al.}(2012){Bernard}, {Ferguson}, {Barker}, {Hidalgo},
  {Ibata}, {Irwin}, {Lewis}, {McConnachie}, {Monelli}, \&
  {Chapman}}]{2012MNRAS.420.2625B}
{Bernard}, E.~J., {Ferguson}, A.~M.~N., {Barker}, M.~K., {et~al.} 2012, \mnras,
  420, 2625

\bibitem[{{Bernard} {et~al.}(2015){Bernard}, {Ferguson}, {Richardson}, {Irwin},
  {Barker}, {Hidalgo}, {Aparicio}, {Chapman}, {Ibata}, {Lewis}, {McConnachie},
  \& {Tanvir}}]{2015MNRAS.446.2789B}
{Bernard}, E.~J., {Ferguson}, A.~M.~N., {Richardson}, J.~C., {et~al.} 2015,
  \mnras, 446, 2789

\bibitem[{{Bica}(1988)}]{1988A&A...195...76B}
{Bica}, E. 1988, \aap, 195, 76

\bibitem[{{Bica} \& {Alloin}(1986{\natexlab{a}})}]{1986A&A...162...21B}
{Bica}, E. \& {Alloin}, D. 1986{\natexlab{a}}, \aap, 162, 21

\bibitem[{{Bica} \& {Alloin}(1986{\natexlab{b}})}]{1986A&AS...66..171B}
{Bica}, E. \& {Alloin}, D. 1986{\natexlab{b}}, \aaps, 66, 171

\bibitem[{{Bica} {et~al.}(1994){Bica}, {Alloin}, \&
  {Schmitt}}]{1994A&A...283..805B}
{Bica}, E., {Alloin}, D., \& {Schmitt}, H.~R. 1994, \aap, 283, 805

\bibitem[{{Bird} {et~al.}(2012){Bird}, {Kazantzidis}, \&
  {Weinberg}}]{2012MNRAS.420..913B}
{Bird}, J.~C., {Kazantzidis}, S., \& {Weinberg}, D.~H. 2012, \mnras, 420, 913

\bibitem[{{Bland-Hawthorn} {et~al.}(2005){Bland-Hawthorn}, {Vlaji{\'c}},
  {Freeman}, \& {Draine}}]{2005ApJ...629..239B}
{Bland-Hawthorn}, J., {Vlaji{\'c}}, M., {Freeman}, K.~C., \& {Draine}, B.~T.
  2005, \apj, 629, 239

\bibitem[{{Cappellari} \& {Copin}(2003)}]{2003MNRAS.342..345C}
{Cappellari}, M. \& {Copin}, Y. 2003, \mnras, 342, 345

\bibitem[{{Cappellari} \& {Emsellem}(2004)}]{2004PASP..116..138C}
{Cappellari}, M. \& {Emsellem}, E. 2004, \pasp, 116, 138

\bibitem[{{Cappellari} {et~al.}(2011){Cappellari}, {Emsellem}, {Krajnovi{\'c}},
  {McDermid}, {Scott}, {Verdoes Kleijn}, {Young}, {Alatalo}, {Bacon}, {Blitz},
  {Bois}, {Bournaud}, {Bureau}, {Davies}, {Davis}, {de Zeeuw}, {Duc},
  {Khochfar}, {Kuntschner}, {Lablanche}, {Morganti}, {Naab}, {Oosterloo},
  {Sarzi}, {Serra}, \& {Weijmans}}]{2011MNRAS.413..813C}
{Cappellari}, M., {Emsellem}, E., {Krajnovi{\'c}}, D., {et~al.} 2011, \mnras,
  413, 813

\bibitem[{{Cid Fernandes} {et~al.}(2005){Cid Fernandes}, {Mateus}, {Sodr{\'e}},
  {Stasi{\'n}ska}, \& {Gomes}}]{2005MNRAS.358..363C}
{Cid Fernandes}, R., {Mateus}, A., {Sodr{\'e}}, L., {Stasi{\'n}ska}, G., \&
  {Gomes}, J.~M. 2005, \mnras, 358, 363

\bibitem[{{Cid Fernandes} {et~al.}(2013){Cid Fernandes}, {P{\'e}rez},
  {Garc{\'{\i}}a Benito}, {Gonz{\'a}lez Delgado}, {de Amorim}, {S{\'a}nchez},
  {Husemann}, {Falc{\'o}n Barroso}, {S{\'a}nchez-Bl{\'a}zquez}, {Walcher}, \&
  {Mast}}]{2013A&A...557A..86C}
{Cid Fernandes}, R., {P{\'e}rez}, E., {Garc{\'{\i}}a Benito}, R., {et~al.}
  2013, \aap, 557, A86

\bibitem[{{Cole} {et~al.}(2007){Cole}, {Skillman}, {Tolstoy}, {Gallagher},
  {Aparicio}, {Dolphin}, {Gallart}, {Hidalgo}, {Saha}, {Stetson}, \&
  {Weisz}}]{2007ApJ...659L..17C}
{Cole}, A.~A., {Skillman}, E.~D., {Tolstoy}, E., {et~al.} 2007, \apjl, 659, L17

\bibitem[{{Dale} {et~al.}(2016){Dale}, {Beltz-Mohrmann}, {Egan}, {Hatlestad},
  {Herzog}, {Leung}, {McLane}, {Phenicie}, {Roberts}, {Barnes}, {Boquien},
  {Calzetti}, {Cook}, {Kobulnicky}, {Staudaher}, \& {van
  Zee}}]{2016AJ....151....4D}
{Dale}, D.~A., {Beltz-Mohrmann}, G.~D., {Egan}, A.~A., {et~al.} 2016, \aj, 151,
  4

\bibitem[{{de Jong}(1996)}]{1996A&A...313..377D}
{de Jong}, R.~S. 1996, \aap, 313, 377

\bibitem[{{Debattista} {et~al.}(2006){Debattista}, {Mayer}, {Carollo}, {Moore},
  {Wadsley}, \& {Quinn}}]{2006ApJ...645..209D}
{Debattista}, V.~P., {Mayer}, L., {Carollo}, C.~M., {et~al.} 2006, \apj, 645,
  209

\bibitem[{{Elmegreen} \& {Hunter}(2006)}]{2006ApJ...636..712E}
{Elmegreen}, B.~G. \& {Hunter}, D.~A. 2006, \apj, 636, 712

\bibitem[{{Erwin} {et~al.}(2008){Erwin}, {Pohlen}, \&
  {Beckman}}]{2008AJ....135...20E}
{Erwin}, P., {Pohlen}, M., \& {Beckman}, J.~E. 2008, \aj, 135, 20

\bibitem[{{Faber} {et~al.}(1985){Faber}, {Friel}, {Burstein}, \&
  {Gaskell}}]{1985ApJS...57..711F}
{Faber}, S.~M., {Friel}, E.~D., {Burstein}, D., \& {Gaskell}, C.~M. 1985,
  \apjs, 57, 711

\bibitem[{{Falc{\'o}n-Barroso} {et~al.}(2006){Falc{\'o}n-Barroso}, {Bacon},
  {Bureau}, {Cappellari}, {Davies}, {de Zeeuw}, {Emsellem}, {Fathi},
  {Krajnovi{\'c}}, {Kuntschner}, {McDermid}, {Peletier}, \&
  {Sarzi}}]{2006MNRAS.369..529F}
{Falc{\'o}n-Barroso}, J., {Bacon}, R., {Bureau}, M., {et~al.} 2006, \mnras,
  369, 529

\bibitem[{{Falc{\'o}n-Barroso} {et~al.}(2017){Falc{\'o}n-Barroso}, {Lyubenova},
  {van de Ven}, {Mendez-Abreu}, {Aguerri}, {Garc{\'{\i}}a-Lorenzo},
  {Bekerait{\'e}}, {S{\'a}nchez}, {Husemann}, {Garc{\'{\i}}a-Benito}, {Mast},
  {Walcher}, {Zibetti}, {Barrera-Ballesteros}, {Galbany},
  {S{\'a}nchez-Bl{\'a}zquez}, {Singh}, {van den Bosch}, {Wild}, {Zhu},
  {Bland-Hawthorn}, {Cid Fernandes}, {de Lorenzo-C{\'a}ceres}, {Gallazzi},
  {Gonz{\'a}lez Delgado}, {Marino}, {M{\'a}rquez}, {P{\'e}rez}, {P{\'e}rez},
  {Roth}, {Rosales-Ortega}, {Ruiz-Lara}, {Wisotzki}, {Ziegler}, \& {Califa
  Collaboration}}]{2017A&A...597A..48F}
{Falc{\'o}n-Barroso}, J., {Lyubenova}, M., {van de Ven}, G., {et~al.} 2017,
  \aap, 597, A48

\bibitem[{{Falc{\'o}n-Barroso} {et~al.}(2011){Falc{\'o}n-Barroso},
  {S{\'a}nchez-Bl{\'a}zquez}, {Vazdekis}, {Ricciardelli}, {Cardiel}, {Cenarro},
  {Gorgas}, \& {Peletier}}]{2011A&A...532A..95F}
{Falc{\'o}n-Barroso}, J., {S{\'a}nchez-Bl{\'a}zquez}, P., {Vazdekis}, A.,
  {et~al.} 2011, \aap, 532, A95

\bibitem[{{Feltzing} {et~al.}(2001){Feltzing}, {Holmberg}, \&
  {Hurley}}]{2001A&A...377..911F}
{Feltzing}, S., {Holmberg}, J., \& {Hurley}, J.~R. 2001, \aap, 377, 911

\bibitem[{{Few} {et~al.}(2012){Few}, {Gibson}, {Courty}, {Michel-Dansac},
  {Brook}, \& {Stinson}}]{2012A&A...547A..63F}
{Few}, C.~G., {Gibson}, B.~K., {Courty}, S., {et~al.} 2012, \aap, 547, A63

\bibitem[{{Freeman}(1970)}]{1970ApJ...160..811F}
{Freeman}, K.~C. 1970, \apj, 160, 811

\bibitem[{{Gallart} {et~al.}(2015){Gallart}, {Monelli}, {Mayer}, {Aparicio},
  {Battaglia}, {Bernard}, {Cassisi}, {Cole}, {Dolphin}, {Drozdovsky},
  {Hidalgo}, {Navarro}, {Salvadori}, {Skillman}, {Stetson}, \&
  {Weisz}}]{2015ApJ...811L..18G}
{Gallart}, C., {Monelli}, M., {Mayer}, L., {et~al.} 2015, \apjl, 811, L18

\bibitem[{{Gallazzi} {et~al.}(2008){Gallazzi}, {Brinchmann}, {Charlot}, \&
  {White}}]{2008MNRAS.383.1439G}
{Gallazzi}, A., {Brinchmann}, J., {Charlot}, S., \& {White}, S.~D.~M. 2008,
  \mnras, 383, 1439

\bibitem[{{Gallazzi} {et~al.}(2006){Gallazzi}, {Charlot}, {Brinchmann}, \&
  {White}}]{2006MNRAS.370.1106G}
{Gallazzi}, A., {Charlot}, S., {Brinchmann}, J., \& {White}, S.~D.~M. 2006,
  \mnras, 370, 1106

\bibitem[{{Gallazzi} {et~al.}(2005){Gallazzi}, {Charlot}, {Brinchmann},
  {White}, \& {Tremonti}}]{2005MNRAS.362...41G}
{Gallazzi}, A., {Charlot}, S., {Brinchmann}, J., {White}, S.~D.~M., \&
  {Tremonti}, C.~A. 2005, \mnras, 362, 41

\bibitem[{{Ganda} {et~al.}(2009){Ganda}, {Peletier}, {Balcells}, \&
  {Falc{\'o}n-Barroso}}]{2009MNRAS.395.1669G}
{Ganda}, K., {Peletier}, R.~F., {Balcells}, M., \& {Falc{\'o}n-Barroso}, J.
  2009, \mnras, 395, 1669

\bibitem[{{Garc{\'{\i}}a-Benito} \&
  {P{\'e}rez-Montero}(2012)}]{2012MNRAS.423..406G}
{Garc{\'{\i}}a-Benito}, R. \& {P{\'e}rez-Montero}, E. 2012, \mnras, 423, 406

\bibitem[{{Garc{\'{\i}}a-Benito} {et~al.}(2015){Garc{\'{\i}}a-Benito},
  {Zibetti}, {S{\'a}nchez}, {Husemann}, {de Amorim}, {Castillo-Morales}, {Cid
  Fernandes}, {Ellis}, {Falc{\'o}n-Barroso}, {Galbany}, {Gil de Paz},
  {Gonz{\'a}lez Delgado}, {Lacerda}, {L{\'o}pez-Fernandez}, {de
  Lorenzo-C{\'a}ceres}, {Lyubenova}, {Marino}, {Mast}, {Mendoza}, {P{\'e}rez},
  {Vale Asari}, {Aguerri}, {Ascasibar}, {Bekerait*error*{\.e}},
  {Bland-Hawthorn}, {Barrera-Ballesteros}, {Bomans}, {Cano-D{\'{\i}}az},
  {Catal{\'a}n-Torrecilla}, {Cortijo}, {Delgado-Inglada}, {Demleitner},
  {Dettmar}, {D{\'{\i}}az}, {Florido}, {Gallazzi}, {Garc{\'{\i}}a-Lorenzo},
  {Gomes}, {Holmes}, {Iglesias-P{\'a}ramo}, {Jahnke}, {Kalinova}, {Kehrig},
  {Kennicutt}, {L{\'o}pez-S{\'a}nchez}, {M{\'a}rquez}, {Masegosa}, {Meidt},
  {Mendez-Abreu}, {Moll{\'a}}, {Monreal-Ibero}, {Morisset}, {del Olmo},
  {Papaderos}, {P{\'e}rez}, {Quirrenbach}, {Rosales-Ortega}, {Roth},
  {Ruiz-Lara}, {S{\'a}nchez-Bl{\'a}zquez}, {S{\'a}nchez-Menguiano}, {Singh},
  {Spekkens}, {Stanishev}, {Torres-Papaqui}, {van de Ven}, {Vilchez},
  {Walcher}, {Wild}, {Wisotzki}, {Ziegler}, {Alves}, {Barrado}, {Quintana}, \&
  {Aceituno}}]{2015A&A...576A.135G}
{Garc{\'{\i}}a-Benito}, R., {Zibetti}, S., {S{\'a}nchez}, S.~F., {et~al.} 2015,
  \aap, 576, A135

\bibitem[{{Gerhard}(1993)}]{1993MNRAS.265..213G}
{Gerhard}, O.~E. 1993, \mnras, 265, 213

\bibitem[{{Gonz{\'a}lez Delgado} {et~al.}(2016){Gonz{\'a}lez Delgado}, {Cid
  Fernandes}, {P{\'e}rez}, {Garc{\'{\i}}a-Benito}, {L{\'o}pez Fern{\'a}ndez},
  {Lacerda}, {Cortijo-Ferrero}, {de Amorim}, {Vale Asari}, {S{\'a}nchez},
  {Walcher}, {Wisotzki}, {Mast}, {Alves}, {Ascasibar}, {Bland-Hawthorn},
  {Galbany}, {Kennicutt}, {M{\'a}rquez}, {Masegosa}, {Moll{\'a}},
  {S{\'a}nchez-Bl{\'a}zquez}, \& {V{\'{\i}}lchez}}]{2016A&A...590A..44G}
{Gonz{\'a}lez Delgado}, R.~M., {Cid Fernandes}, R., {P{\'e}rez}, E., {et~al.}
  2016, \aap, 590, A44

\bibitem[{{Gonz{\'a}lez Delgado} {et~al.}(2015){Gonz{\'a}lez Delgado},
  {Garc{\'{\i}}a-Benito}, {P{\'e}rez}, {Cid Fernandes}, {de Amorim},
  {Cortijo-Ferrero}, {Lacerda}, {L{\'o}pez Fern{\'a}ndez}, {Vale-Asari},
  {S{\'a}nchez}, {Moll{\'a}}, {Ruiz-Lara}, {S{\'a}nchez-Bl{\'a}zquez},
  {Walcher}, {Alves}, {Aguerri}, {Bekerait{\'e}}, {Bland-Hawthorn}, {Galbany},
  {Gallazzi}, {Husemann}, {Iglesias-P{\'a}ramo}, {Kalinova},
  {L{\'o}pez-S{\'a}nchez}, {Marino}, {M{\'a}rquez}, {Masegosa}, {Mast},
  {M{\'e}ndez-Abreu}, {Mendoza}, {del Olmo}, {P{\'e}rez}, {Quirrenbach}, \&
  {Zibetti}}]{2015A&A...581A.103G}
{Gonz{\'a}lez Delgado}, R.~M., {Garc{\'{\i}}a-Benito}, R., {P{\'e}rez}, E.,
  {et~al.} 2015, \aap, 581, A103

\bibitem[{{Gorgas} {et~al.}(1993){Gorgas}, {Faber}, {Burstein}, {Gonzalez},
  {Courteau}, \& {Prosser}}]{1993ApJS...86..153G}
{Gorgas}, J., {Faber}, S.~M., {Burstein}, D., {et~al.} 1993, \apjs, 86, 153

\bibitem[{{Guti{\'e}rrez} {et~al.}(2011){Guti{\'e}rrez}, {Erwin}, {Aladro}, \&
  {Beckman}}]{2011AJ....142..145G}
{Guti{\'e}rrez}, L., {Erwin}, P., {Aladro}, R., \& {Beckman}, J.~E. 2011, \aj,
  142, 145

\bibitem[{{Heavens} {et~al.}(2000){Heavens}, {Jimenez}, \&
  {Lahav}}]{2000MNRAS.317..965H}
{Heavens}, A.~F., {Jimenez}, R., \& {Lahav}, O. 2000, \mnras, 317, 965

\bibitem[{{Hidalgo} {et~al.}(2011){Hidalgo}, {Aparicio}, {Skillman}, {Monelli},
  {Gallart}, {Cole}, {Dolphin}, {Weisz}, {Bernard}, {Cassisi}, {Mayer},
  {Stetson}, {Tolstoy}, \& {Ferguson}}]{2011ApJ...730...14H}
{Hidalgo}, S.~L., {Aparicio}, A., {Skillman}, E., {et~al.} 2011, \apj, 730, 14

\bibitem[{{Husemann} {et~al.}(2013){Husemann}, {Jahnke}, {S{\'a}nchez},
  {Barrado}, {Bekerait*error*{\.e}}, {Bomans}, {Castillo-Morales},
  {Catal{\'a}n-Torrecilla}, {Cid Fernandes}, {Falc{\'o}n-Barroso},
  {Garc{\'{\i}}a-Benito}, {Gonz{\'a}lez Delgado}, {Iglesias-P{\'a}ramo},
  {Johnson}, {Kupko}, {L{\'o}pez-Fernandez}, {Lyubenova}, {Marino}, {Mast},
  {Miskolczi}, {Monreal-Ibero}, {Gil de Paz}, {P{\'e}rez}, {P{\'e}rez},
  {Rosales-Ortega}, {Ruiz-Lara}, {Schilling}, {van de Ven}, {Walcher}, {Alves},
  {de Amorim}, {Backsmann}, {Barrera-Ballesteros}, {Bland-Hawthorn}, {Cortijo},
  {Dettmar}, {Demleitner}, {D{\'{\i}}az}, {Enke}, {Florido}, {Flores},
  {Galbany}, {Gallazzi}, {Garc{\'{\i}}a-Lorenzo}, {Gomes}, {Gruel}, {Haines},
  {Holmes}, {Jungwiert}, {Kalinova}, {Kehrig}, {Kennicutt}, {Klar}, {Lehnert},
  {L{\'o}pez-S{\'a}nchez}, {de Lorenzo-C{\'a}ceres}, {M{\'a}rmol-Queralt{\'o}},
  {M{\'a}rquez}, {Mendez-Abreu}, {Moll{\'a}}, {del Olmo}, {Meidt}, {Papaderos},
  {Puschnig}, {Quirrenbach}, {Roth}, {S{\'a}nchez-Bl{\'a}zquez}, {Spekkens},
  {Singh}, {Stanishev}, {Trager}, {Vilchez}, {Wild}, {Wisotzki}, {Zibetti}, \&
  {Ziegler}}]{2013A&A...549A..87H}
{Husemann}, B., {Jahnke}, K., {S{\'a}nchez}, S.~F., {et~al.} 2013, \aap, 549,
  A87

\bibitem[{{Jansen} {et~al.}(2000){Jansen}, {Fabricant}, {Franx}, \&
  {Caldwell}}]{2000ApJS..126..331J}
{Jansen}, R.~A., {Fabricant}, D., {Franx}, M., \& {Caldwell}, N. 2000, \apjs,
  126, 331

\bibitem[{{Kauffmann} {et~al.}(2003){Kauffmann}, {Heckman}, {White}, {Charlot},
  {Tremonti}, {Brinchmann}, {Bruzual}, {Peng}, {Seibert}, {Bernardi},
  {Blanton}, {Brinkmann}, {Castander}, {Cs{\'a}bai}, {Fukugita}, {Ivezic},
  {Munn}, {Nichol}, {Padmanabhan}, {Thakar}, {Weinberg}, \&
  {York}}]{2003MNRAS.341...33K}
{Kauffmann}, G., {Heckman}, T.~M., {White}, S.~D.~M., {et~al.} 2003, \mnras,
  341, 33

\bibitem[{{Koleva} {et~al.}(2009){Koleva}, {Prugniel}, {Bouchard}, \&
  {Wu}}]{2009A&A...501.1269K}
{Koleva}, M., {Prugniel}, P., {Bouchard}, A., \& {Wu}, Y. 2009, \aap, 501, 1269

\bibitem[{{Kordopatis} {et~al.}(2015){Kordopatis}, {Binney}, {Gilmore}, {Wyse},
  {Belokurov}, {McMillan}, {Hatfield}, {Grebel}, {Steinmetz}, {Navarro},
  {Seabroke}, {Minchev}, {Chiappini}, {Bienaym{\'e}}, {Bland-Hawthorn},
  {Freeman}, {Gibson}, {Helmi}, {Munari}, {Parker}, {Reid}, {Siebert},
  {Siviero}, \& {Zwitter}}]{2015MNRAS.447.3526K}
{Kordopatis}, G., {Binney}, J., {Gilmore}, G., {et~al.} 2015, \mnras, 447, 3526

\bibitem[{{Kordopatis} {et~al.}(2013){Kordopatis}, {Gilmore}, {Steinmetz},
  {Boeche}, {Seabroke}, {Siebert}, {Zwitter}, {Binney}, {de Laverny},
  {Recio-Blanco}, {Williams}, {Piffl}, {Enke}, {Roeser}, {Bijaoui}, {Wyse},
  {Freeman}, {Munari}, {Carrillo}, {Anguiano}, {Burton}, {Campbell}, {Cass},
  {Fiegert}, {Hartley}, {Parker}, {Reid}, {Ritter}, {Russell}, {Stupar},
  {Watson}, {Bienaym{\'e}}, {Bland-Hawthorn}, {Gerhard}, {Gibson}, {Grebel},
  {Helmi}, {Navarro}, {Conrad}, {Famaey}, {Faure}, {Just}, {Kos}, {Matijevi{\v
  c}}, {McMillan}, {Minchev}, {Scholz}, {Sharma}, {Siviero}, {de Boer}, \& {{\v
  Z}erjal}}]{2013AJ....146..134K}
{Kordopatis}, G., {Gilmore}, G., {Steinmetz}, M., {et~al.} 2013, \aj, 146, 134

\bibitem[{{Kroupa}(2001)}]{2001MNRAS.322..231K}
{Kroupa}, P. 2001, \mnras, 322, 231

\bibitem[{{Kuncarayakti} {et~al.}(2016){Kuncarayakti}, {Galbany}, {Anderson},
  {Kr{\"u}hler}, \& {Hamuy}}]{2016A&A...593A..78K}
{Kuncarayakti}, H., {Galbany}, L., {Anderson}, J.~P., {Kr{\"u}hler}, T., \&
  {Hamuy}, M. 2016, \aap, 593, A78

\bibitem[{{Kuntschner} {et~al.}(2010){Kuntschner}, {Emsellem}, {Bacon},
  {Cappellari}, {Davies}, {de Zeeuw}, {Falc{\'o}n-Barroso}, {Krajnovi{\'c}},
  {McDermid}, {Peletier}, {Sarzi}, {Shapiro}, {van den Bosch}, \& {van de
  Ven}}]{2010MNRAS.408...97K}
{Kuntschner}, H., {Emsellem}, E., {Bacon}, R., {et~al.} 2010, \mnras, 408, 97

\bibitem[{{MacArthur} {et~al.}(2004){MacArthur}, {Courteau}, {Bell}, \&
  {Holtzman}}]{2004ApJS..152..175M}
{MacArthur}, L.~A., {Courteau}, S., {Bell}, E., \& {Holtzman}, J.~A. 2004,
  \apjs, 152, 175

\bibitem[{{MacArthur} {et~al.}(2009){MacArthur}, {Gonz{\'a}lez}, \&
  {Courteau}}]{2009MNRAS.395...28M}
{MacArthur}, L.~A., {Gonz{\'a}lez}, J.~J., \& {Courteau}, S. 2009, \mnras, 395,
  28

\bibitem[{{Marino} {et~al.}(2016){Marino}, {Gil de Paz}, {S{\'a}nchez},
  {S{\'a}nchez-Bl{\'a}zquez}, {Cardiel}, {Castillo-Morales}, {Pascual},
  {V{\'{\i}}lchez}, {Kehrig}, {Moll{\'a}}, {Mendez-Abreu},
  {Catal{\'a}n-Torrecilla}, {Florido}, {Perez}, {Ruiz-Lara}, {Ellis},
  {L{\'o}pez-S{\'a}nchez}, {Gonz{\'a}lez Delgado}, {de Lorenzo-C{\'a}ceres},
  {Garc{\'{\i}}a-Benito}, {Galbany}, {Zibetti}, {Cortijo}, {Kalinova}, {Mast},
  {Iglesias-P{\'a}ramo}, {Papaderos}, {Walcher}, \&
  {Bland-Hawthorn}}]{2016A&A...585A..47M}
{Marino}, R.~A., {Gil de Paz}, A., {S{\'a}nchez}, S.~F., {et~al.} 2016, \aap,
  585, A47

\bibitem[{{Markwardt}(2009)}]{2009ASPC..411..251M}
{Markwardt}, C.~B. 2009, in Astronomical Society of the Pacific Conference
  Series, Vol. 411, Astronomical Data Analysis Software and Systems XVIII, ed.
  D.~A. {Bohlender}, D.~{Durand}, \& P.~{Dowler}, 251

\bibitem[{{Mart{\'{\i}}nez-Serrano} {et~al.}(2009){Mart{\'{\i}}nez-Serrano},
  {Serna}, {Dom{\'e}nech-Moral}, \&
  {Dom{\'{\i}}nguez-Tenreiro}}]{2009ApJ...705L.133M}
{Mart{\'{\i}}nez-Serrano}, F.~J., {Serna}, A., {Dom{\'e}nech-Moral}, M., \&
  {Dom{\'{\i}}nguez-Tenreiro}, R. 2009, \apjl, 705, L133

\bibitem[{{M{\'e}ndez-Abreu} {et~al.}(2008){M{\'e}ndez-Abreu}, {Aguerri},
  {Corsini}, \& {Simonneau}}]{2008A&A...478..353M}
{M{\'e}ndez-Abreu}, J., {Aguerri}, J.~A.~L., {Corsini}, E.~M., \& {Simonneau},
  E. 2008, \aap, 478, 353

\bibitem[{{M{\'e}ndez-Abreu} {et~al.}(2014){M{\'e}ndez-Abreu}, {Debattista},
  {Corsini}, \& {Aguerri}}]{2014A&A...572A..25M}
{M{\'e}ndez-Abreu}, J., {Debattista}, V.~P., {Corsini}, E.~M., \& {Aguerri},
  J.~A.~L. 2014, \aap, 572, A25

\bibitem[{{M{\'e}ndez-Abreu} {et~al.}(2017){M{\'e}ndez-Abreu}, {Ruiz-Lara},
  {S{\'a}nchez-Menguiano}, {de Lorenzo-C{\'a}ceres}, {Costantin},
  {Catal{\'a}n-Torrecilla}, {Florido}, {Aguerri}, {Bland-Hawthorn}, {Corsini},
  {Dettmar}, {Galbany}, {Garc{\'{\i}}a-Benito}, {Marino}, {M{\'a}rquez},
  {Ortega-Minakata}, {Papaderos}, {S{\'a}nchez}, {S{\'a}nchez-Blazquez},
  {Spekkens}, {van de Ven}, {Wild}, \& {Ziegler}}]{2017A&A...598A..32M}
{M{\'e}ndez-Abreu}, J., {Ruiz-Lara}, T., {S{\'a}nchez-Menguiano}, L., {et~al.}
  2017, \aap, 598, A32

\bibitem[{{Minchev} \& {Famaey}(2010)}]{2010ApJ...722..112M}
{Minchev}, I. \& {Famaey}, B. 2010, \apj, 722, 112

\bibitem[{{Minchev} {et~al.}(2011){Minchev}, {Famaey}, {Combes}, {Di Matteo},
  {Mouhcine}, \& {Wozniak}}]{2011A&A...527A.147M}
{Minchev}, I., {Famaey}, B., {Combes}, F., {et~al.} 2011, \aap, 527, A147

\bibitem[{{Minchev} {et~al.}(2012{\natexlab{a}}){Minchev}, {Famaey}, {Quillen},
  {Dehnen}, {Martig}, \& {Siebert}}]{2012A&A...548A.127M}
{Minchev}, I., {Famaey}, B., {Quillen}, A.~C., {et~al.} 2012{\natexlab{a}},
  \aap, 548, A127

\bibitem[{{Minchev} {et~al.}(2012{\natexlab{b}}){Minchev}, {Famaey}, {Quillen},
  {Di Matteo}, {Combes}, {Vlaji{\'c}}, {Erwin}, \&
  {Bland-Hawthorn}}]{2012A&A...548A.126M}
{Minchev}, I., {Famaey}, B., {Quillen}, A.~C., {et~al.} 2012{\natexlab{b}},
  \aap, 548, A126

\bibitem[{{Monelli} {et~al.}(2010){Monelli}, {Hidalgo}, {Stetson}, {Aparicio},
  {Gallart}, {Dolphin}, {Cole}, {Weisz}, {Skillman}, {Bernard}, {Mayer},
  {Navarro}, {Cassisi}, {Drozdovsky}, \& {Tolstoy}}]{2010ApJ...720.1225M}
{Monelli}, M., {Hidalgo}, S.~L., {Stetson}, P.~B., {et~al.} 2010, \apj, 720,
  1225

\bibitem[{{Mor\'e} {et~al.}(1980){Mor\'e}, {Garbow}, \& {Hillstrom}}]{more80}
{Mor\'e}, J.~J., {Garbow}, B.~S., \& {Hillstrom}, K.~E. 1980, Argonne National
  Laboratory Report ANL-80-74 [\eprint[arXiv]{0806.2988}]

\bibitem[{{Morelli} {et~al.}(2015){Morelli}, {Corsini}, {Pizzella}, {Dalla
  Bont{\`a}}, {Coccato}, \& {M{\'e}ndez-Abreu}}]{2015MNRAS.452.1128M}
{Morelli}, L., {Corsini}, E.~M., {Pizzella}, A., {et~al.} 2015, \mnras, 452,
  1128

\bibitem[{{Morelli} {et~al.}(2016){Morelli}, {Parmiggiani}, {Corsini},
  {Costantin}, {Dalla Bont{\`a}}, {M{\'e}ndez-Abreu}, \&
  {Pizzella}}]{2016MNRAS.463.4396M}
{Morelli}, L., {Parmiggiani}, M., {Corsini}, E.~M., {et~al.} 2016, \mnras, 463,
  4396

\bibitem[{{Mu{\~n}oz-Mateos} {et~al.}(2009){Mu{\~n}oz-Mateos}, {Gil de Paz},
  {Zamorano}, {Boissier}, {Dale}, {P{\'e}rez-Gonz{\'a}lez}, {Gallego},
  {Madore}, {Bendo}, {Boselli}, {Buat}, {Calzetti}, {Moustakas}, \&
  {Kennicutt}}]{2009ApJ...703.1569M}
{Mu{\~n}oz-Mateos}, J.~C., {Gil de Paz}, A., {Zamorano}, J., {et~al.} 2009,
  \apj, 703, 1569

\bibitem[{{Nordstr{\"o}m} {et~al.}(2004){Nordstr{\"o}m}, {Mayor}, {Andersen},
  {Holmberg}, {Pont}, {J{\o}rgensen}, {Olsen}, {Udry}, \&
  {Mowlavi}}]{2004A&A...418..989N}
{Nordstr{\"o}m}, B., {Mayor}, M., {Andersen}, J., {et~al.} 2004, \aap, 418, 989

\bibitem[{{Ocvirk}(2010)}]{2010ApJ...709...88O}
{Ocvirk}, P. 2010, \apj, 709, 88

\bibitem[{{Ocvirk} {et~al.}(2006{\natexlab{a}}){Ocvirk}, {Pichon}, {Lan{\c
  c}on}, \& {Thi{\'e}baut}}]{2006MNRAS.365...74O}
{Ocvirk}, P., {Pichon}, C., {Lan{\c c}on}, A., \& {Thi{\'e}baut}, E.
  2006{\natexlab{a}}, \mnras, 365, 74

\bibitem[{{Ocvirk} {et~al.}(2006{\natexlab{b}}){Ocvirk}, {Pichon}, {Lan{\c
  c}on}, \& {Thi{\'e}baut}}]{2006MNRAS.365...46O}
{Ocvirk}, P., {Pichon}, C., {Lan{\c c}on}, A., \& {Thi{\'e}baut}, E.
  2006{\natexlab{b}}, \mnras, 365, 46

\bibitem[{{Peletier}(1993)}]{1993A&A...271...51P}
{Peletier}, R.~F. 1993, \aap, 271, 51

\bibitem[{{Peletier} \& {Balcells}(1996)}]{1996AJ....111.2238P}
{Peletier}, R.~F. \& {Balcells}, M. 1996, \aj, 111, 2238

\bibitem[{{Peletier} {et~al.}(2007){Peletier}, {Falc{\'o}n-Barroso}, {Bacon},
  {Cappellari}, {Davies}, {de Zeeuw}, {Emsellem}, {Ganda}, {Krajnovi{\'c}},
  {Kuntschner}, {McDermid}, {Sarzi}, \& {van de Ven}}]{2007MNRAS.379..445P}
{Peletier}, R.~F., {Falc{\'o}n-Barroso}, J., {Bacon}, R., {et~al.} 2007,
  \mnras, 379, 445

\bibitem[{{P{\'e}rez} {et~al.}(2013){P{\'e}rez}, {Cid Fernandes}, {Gonz{\'a}lez
  Delgado}, {Garc{\'{\i}}a-Benito}, {S{\'a}nchez}, {Husemann}, {Mast},
  {Rod{\'o}n}, {Kupko}, {Backsmann}, {de Amorim}, {van de Ven}, {Walcher},
  {Wisotzki}, {Cortijo-Ferrero}, \& {CALIFA
  Collaboration}}]{2013ApJ...764L...1P}
{P{\'e}rez}, E., {Cid Fernandes}, R., {Gonz{\'a}lez Delgado}, R.~M., {et~al.}
  2013, \apjl, 764, L1

\bibitem[{{P{\'e}rez} \&
  {S{\'a}nchez-Bl{\'a}zquez}(2011)}]{2011A&A...529A..64P}
{P{\'e}rez}, I. \& {S{\'a}nchez-Bl{\'a}zquez}, P. 2011, \aap, 529, A64

\bibitem[{{Pohlen} \& {Trujillo}(2006)}]{2006A&A...454..759P}
{Pohlen}, M. \& {Trujillo}, I. 2006, \aap, 454, 759

\bibitem[{{Radburn-Smith} {et~al.}(2012){Radburn-Smith}, {Ro{\v s}kar},
  {Debattista}, {Dalcanton}, {Streich}, {de Jong}, {Vlaji{\'c}}, {Holwerda},
  {Purcell}, {Dolphin}, \& {Zucker}}]{2012ApJ...753..138R}
{Radburn-Smith}, D.~J., {Ro{\v s}kar}, R., {Debattista}, V.~P., {et~al.} 2012,
  \apj, 753, 138

\bibitem[{{Reichardt} {et~al.}(2001){Reichardt}, {Jimenez}, \&
  {Heavens}}]{2001MNRAS.327..849R}
{Reichardt}, C., {Jimenez}, R., \& {Heavens}, A.~F. 2001, \mnras, 327, 849

\bibitem[{{Roediger} {et~al.}(2011{\natexlab{a}}){Roediger}, {Courteau},
  {MacArthur}, \& {McDonald}}]{2011MNRAS.416.1996R}
{Roediger}, J.~C., {Courteau}, S., {MacArthur}, L.~A., \& {McDonald}, M.
  2011{\natexlab{a}}, \mnras, 416, 1996

\bibitem[{{Roediger} {et~al.}(2011{\natexlab{b}}){Roediger}, {Courteau},
  {McDonald}, \& {MacArthur}}]{2011MNRAS.416.1983R}
{Roediger}, J.~C., {Courteau}, S., {McDonald}, M., \& {MacArthur}, L.~A.
  2011{\natexlab{b}}, \mnras, 416, 1983

\bibitem[{{Roediger} {et~al.}(2012){Roediger}, {Courteau},
  {S{\'a}nchez-Bl{\'a}zquez}, \& {McDonald}}]{2012ApJ...758...41R}
{Roediger}, J.~C., {Courteau}, S., {S{\'a}nchez-Bl{\'a}zquez}, P., \&
  {McDonald}, M. 2012, \apj, 758, 41

\bibitem[{{Rose}(1984)}]{1984AJ.....89.1238R}
{Rose}, J.~A. 1984, \aj, 89, 1238

\bibitem[{{Roth} {et~al.}(2005){Roth}, {Kelz}, {Fechner}, {Hahn}, {Bauer},
  {Becker}, {B{\"o}hm}, {Christensen}, {Dionies}, {Paschke}, {Popow}, {Wolter},
  {Schmoll}, {Laux}, \& {Altmann}}]{2005PASP..117..620R}
{Roth}, M.~M., {Kelz}, A., {Fechner}, T., {et~al.} 2005, \pasp, 117, 620

\bibitem[{{Ro{\v s}kar} {et~al.}(2008{\natexlab{a}}){Ro{\v s}kar},
  {Debattista}, {Quinn}, {Stinson}, \& {Wadsley}}]{2008ApJ...684L..79R}
{Ro{\v s}kar}, R., {Debattista}, V.~P., {Quinn}, T.~R., {Stinson}, G.~S., \&
  {Wadsley}, J. 2008{\natexlab{a}}, \apjl, 684, L79

\bibitem[{{Ro{\v s}kar} {et~al.}(2012){Ro{\v s}kar}, {Debattista}, {Quinn}, \&
  {Wadsley}}]{2012MNRAS.426.2089R}
{Ro{\v s}kar}, R., {Debattista}, V.~P., {Quinn}, T.~R., \& {Wadsley}, J. 2012,
  \mnras, 426, 2089

\bibitem[{{Ro{\v s}kar} {et~al.}(2008{\natexlab{b}}){Ro{\v s}kar},
  {Debattista}, {Stinson}, {Quinn}, {Kaufmann}, \&
  {Wadsley}}]{2008ApJ...675L..65R}
{Ro{\v s}kar}, R., {Debattista}, V.~P., {Stinson}, G.~S., {et~al.}
  2008{\natexlab{b}}, \apjl, 675, L65

\bibitem[{{Ruiz-Lara} {et~al.}(2016{\natexlab{a}}){Ruiz-Lara}, {Few}, {Gibson},
  {P{\'e}rez}, {Florido}, {Minchev}, \&
  {S{\'a}nchez-Bl{\'a}zquez}}]{2016A&A...586A.112R}
{Ruiz-Lara}, T., {Few}, C.~G., {Gibson}, B.~K., {et~al.} 2016{\natexlab{a}},
  \aap, 586, A112

\bibitem[{{Ruiz-Lara} {et~al.}(2016{\natexlab{b}}){Ruiz-Lara}, {P{\'e}rez},
  {Florido}, {S{\'a}nchez-Bl{\'a}zquez}, {M{\'e}ndez-Abreu}, {Lyubenova},
  {Falc{\'o}n-Barroso}, {S{\'a}nchez-Menguiano}, {S{\'a}nchez}, {Galbany},
  {Garc{\'{\i}}a-Benito}, {Gonz{\'a}lez Delgado}, {Husemann}, {Kehrig},
  {L{\'o}pez-S{\'a}nchez}, {Marino}, {Mast}, {Papaderos}, {van de Ven},
  {Walcher}, {Zibetti}, \& {CALIFA Team}}]{2016MNRAS.456L..35R}
{Ruiz-Lara}, T., {P{\'e}rez}, I., {Florido}, E., {et~al.} 2016{\natexlab{b}},
  \mnras, 456, L35

\bibitem[{{Ruiz-Lara} {et~al.}(2015){Ruiz-Lara}, {P{\'e}rez}, {Gallart},
  {Alloin}, {Monelli}, {Koleva}, {Pompei}, {Beasley},
  {S{\'a}nchez-Bl{\'a}zquez}, {Florido}, {Aparicio}, {Fleurence}, {Hardy},
  {Hidalgo}, \& {Raimann}}]{2015A&A...583A..60R}
{Ruiz-Lara}, T., {P{\'e}rez}, I., {Gallart}, C., {et~al.} 2015, \aap, 583, A60

\bibitem[{{S{\'a}nchez} {et~al.}(2016){S{\'a}nchez}, {Garc{\'{\i}}a-Benito},
  {Zibetti}, {Walcher}, {Husemann}, {Mendoza}, {Galbany}, {Falc{\'o}n-Barroso},
  {Mast}, {Aceituno}, {Aguerri}, {Alves}, {Amorim}, {Ascasibar},
  {Barrado-Navascues}, {Barrera-Ballesteros}, {Bekerait{\`e}},
  {Bland-Hawthorn}, {Cano D{\'{\i}}az}, {Cid Fernandes}, {Cavichia}, {Cortijo},
  {Dannerbauer}, {Demleitner}, {D{\'{\i}}az}, {Dettmar}, {de
  Lorenzo-C{\'a}ceres}, {del Olmo}, {Galazzi}, {Garc{\'{\i}}a-Lorenzo}, {Gil de
  Paz}, {Gonz{\'a}lez Delgado}, {Holmes}, {Igl{\'e}sias-P{\'a}ramo}, {Kehrig},
  {Kelz}, {Kennicutt}, {Kleemann}, {Lacerda}, {L{\'o}pez Fern{\'a}ndez},
  {L{\'o}pez S{\'a}nchez}, {Lyubenova}, {Marino}, {M{\'a}rquez},
  {Mendez-Abreu}, {Moll{\'a}}, {Monreal-Ibero}, {Ortega Minakata},
  {Torres-Papaqui}, {P{\'e}rez}, {Rosales-Ortega}, {Roth},
  {S{\'a}nchez-Bl{\'a}zquez}, {Schilling}, {Spekkens}, {Vale Asari}, {van den
  Bosch}, {van de Ven}, {Vilchez}, {Wild}, {Wisotzki}, {Y{\i}ld{\i}r{\i}m}, \&
  {Ziegler}}]{2016A&A...594A..36S}
{S{\'a}nchez}, S.~F., {Garc{\'{\i}}a-Benito}, R., {Zibetti}, S., {et~al.} 2016,
  \aap, 594, A36

\bibitem[{{S{\'a}nchez} {et~al.}(2012){S{\'a}nchez}, {Kennicutt}, {Gil de Paz},
  {van de Ven}, {V{\'{\i}}lchez}, {Wisotzki}, {Walcher}, {Mast}, {Aguerri},
  {Albiol-P{\'e}rez}, {Alonso-Herrero}, {Alves}, {Bakos}, {Bart{\'a}kov{\'a}},
  {Bland-Hawthorn}, {Boselli}, {Bomans}, {Castillo-Morales}, {Cortijo-Ferrero},
  {de Lorenzo-C{\'a}ceres}, {Del Olmo}, {Dettmar}, {D{\'{\i}}az}, {Ellis},
  {Falc{\'o}n-Barroso}, {Flores}, {Gallazzi}, {Garc{\'{\i}}a-Lorenzo},
  {Gonz{\'a}lez Delgado}, {Gruel}, {Haines}, {Hao}, {Husemann},
  {Igl{\'e}sias-P{\'a}ramo}, {Jahnke}, {Johnson}, {Jungwiert}, {Kalinova},
  {Kehrig}, {Kupko}, {L{\'o}pez-S{\'a}nchez}, {Lyubenova}, {Marino},
  {M{\'a}rmol-Queralt{\'o}}, {M{\'a}rquez}, {Masegosa}, {Meidt},
  {Mendez-Abreu}, {Monreal-Ibero}, {Montijo}, {Mour{\~a}o}, {Palacios-Navarro},
  {Papaderos}, {Pasquali}, {Peletier}, {P{\'e}rez}, {P{\'e}rez}, {Quirrenbach},
  {Rela{\~n}o}, {Rosales-Ortega}, {Roth}, {Ruiz-Lara},
  {S{\'a}nchez-Bl{\'a}zquez}, {Sengupta}, {Singh}, {Stanishev}, {Trager},
  {Vazdekis}, {Viironen}, {Wild}, {Zibetti}, \&
  {Ziegler}}]{2012A&A...538A...8S}
{S{\'a}nchez}, S.~F., {Kennicutt}, R.~C., {Gil de Paz}, A., {et~al.} 2012,
  \aap, 538, A8

\bibitem[{{S{\'a}nchez} {et~al.}(2011){S{\'a}nchez}, {Rosales-Ortega},
  {Kennicutt}, {Johnson}, {Diaz}, {Pasquali}, \& {Hao}}]{2011MNRAS.410..313S}
{S{\'a}nchez}, S.~F., {Rosales-Ortega}, F.~F., {Kennicutt}, R.~C., {et~al.}
  2011, \mnras, 410, 313

\bibitem[{{S{\'a}nchez-Bl{\'a}zquez} {et~al.}(2009){S{\'a}nchez-Bl{\'a}zquez},
  {Courty}, {Gibson}, \& {Brook}}]{2009MNRAS.398..591S}
{S{\'a}nchez-Bl{\'a}zquez}, P., {Courty}, S., {Gibson}, B.~K., \& {Brook},
  C.~B. 2009, \mnras, 398, 591

\bibitem[{{S{\'a}nchez-Bl{\'a}zquez} {et~al.}(2011){S{\'a}nchez-Bl{\'a}zquez},
  {Ocvirk}, {Gibson}, {P{\'e}rez}, \& {Peletier}}]{2011MNRAS.415..709S}
{S{\'a}nchez-Bl{\'a}zquez}, P., {Ocvirk}, P., {Gibson}, B.~K., {P{\'e}rez}, I.,
  \& {Peletier}, R.~F. 2011, \mnras, 415, 709

\bibitem[{{S{\'a}nchez-Bl{\'a}zquez} {et~al.}(2006){S{\'a}nchez-Bl{\'a}zquez},
  {Peletier}, {Jim{\'e}nez-Vicente}, {Cardiel}, {Cenarro},
  {Falc{\'o}n-Barroso}, {Gorgas}, {Selam}, \& {Vazdekis}}]{2006MNRAS.371..703S}
{S{\'a}nchez-Bl{\'a}zquez}, P., {Peletier}, R.~F., {Jim{\'e}nez-Vicente}, J.,
  {et~al.} 2006, \mnras, 371, 703

\bibitem[{{S{\'a}nchez-Bl{\'a}zquez}
  {et~al.}(2014{\natexlab{a}}){S{\'a}nchez-Bl{\'a}zquez}, {Rosales-Ortega},
  {Diaz}, \& {S{\'a}nchez}}]{2014MNRAS.437.1534S}
{S{\'a}nchez-Bl{\'a}zquez}, P., {Rosales-Ortega}, F., {Diaz}, A., \&
  {S{\'a}nchez}, S.~F. 2014{\natexlab{a}}, \mnras, 437, 1534

\bibitem[{{S{\'a}nchez-Bl{\'a}zquez}
  {et~al.}(2014{\natexlab{b}}){S{\'a}nchez-Bl{\'a}zquez}, {Rosales-Ortega},
  {M{\'e}ndez-Abreu}, {P{\'e}rez}, {S{\'a}nchez}, {Zibetti}, {Aguerri},
  {Bland-Hawthorn}, {Catal{\'a}n-Torrecilla}, {Cid Fernandes}, {de Amorim}, {de
  Lorenzo-Caceres}, {Falc{\'o}n-Barroso}, {Galazzi}, {Garc{\'{\i}}a Benito},
  {Gil de Paz}, {Gonz{\'a}lez Delgado}, {Husemann}, {Iglesias-P{\'a}ramo},
  {Jungwiert}, {Marino}, {M{\'a}rquez}, {Mast}, {Mendoza}, {Moll{\'a}},
  {Papaderos}, {Ruiz-Lara}, {van de Ven}, {Walcher}, \&
  {Wisotzki}}]{2014A&A...570A...6S}
{S{\'a}nchez-Bl{\'a}zquez}, P., {Rosales-Ortega}, F.~F., {M{\'e}ndez-Abreu},
  J., {et~al.} 2014{\natexlab{b}}, \aap, 570, A6

\bibitem[{{S{\'a}nchez-Menguiano} {et~al.}(2016){S{\'a}nchez-Menguiano},
  {S{\'a}nchez}, {P{\'e}rez}, {Garc{\'{\i}}a-Benito}, {Husemann}, {Mast},
  {Mendoza}, {Ruiz-Lara}, {Ascasibar}, {Bland-Hawthorn}, {Cavichia},
  {D{\'{\i}}az}, {Florido}, {Galbany}, {G{\'o}nzalez Delgado}, {Kehrig},
  {Marino}, {M{\'a}rquez}, {Masegosa}, {M{\'e}ndez-Abreu}, {Moll{\'a}}, {Del
  Olmo}, {P{\'e}rez}, {S{\'a}nchez-Bl{\'a}zquez}, {Stanishev}, {Walcher},
  {L{\'o}pez-S{\'a}nchez}, \& {Califa Collaboration}}]{2016A&A...587A..70S}
{S{\'a}nchez-Menguiano}, L., {S{\'a}nchez}, S.~F., {P{\'e}rez}, I., {et~al.}
  2016, \aap, 587, A70

\bibitem[{{Sarzi} {et~al.}(2006){Sarzi}, {Falc{\'o}n-Barroso}, {Davies},
  {Bacon}, {Bureau}, {Cappellari}, {de Zeeuw}, {Emsellem}, {Fathi},
  {Krajnovi{\'c}}, {Kuntschner}, {McDermid}, \&
  {Peletier}}]{2006MNRAS.366.1151S}
{Sarzi}, M., {Falc{\'o}n-Barroso}, J., {Davies}, R.~L., {et~al.} 2006, \mnras,
  366, 1151

\bibitem[{{Seidel} {et~al.}(2015){Seidel}, {Cacho}, {Ruiz-Lara},
  {Falc{\'o}n-Barroso}, {P{\'e}rez}, {S{\'a}nchez-Bl{\'a}zquez}, {Vogt},
  {Ness}, {Freeman}, \& {Aniyan}}]{2015MNRAS.446.2837S}
{Seidel}, M.~K., {Cacho}, R., {Ruiz-Lara}, T., {et~al.} 2015, \mnras, 446, 2837

\bibitem[{{Sellwood} \& {Binney}(2002)}]{2002MNRAS.336..785S}
{Sellwood}, J.~A. \& {Binney}, J.~J. 2002, \mnras, 336, 785

\bibitem[{{Steinmetz} {et~al.}(2006){Steinmetz}, {Zwitter}, {Siebert},
  {Watson}, {Freeman}, {Munari}, {Campbell}, {Williams}, {Seabroke}, {Wyse},
  {Parker}, {Bienaym{\'e}}, {Roeser}, {Gibson}, {Gilmore}, {Grebel}, {Helmi},
  {Navarro}, {Burton}, {Cass}, {Dawe}, {Fiegert}, {Hartley}, {Russell},
  {Saunders}, {Enke}, {Bailin}, {Binney}, {Bland-Hawthorn}, {Boeche}, {Dehnen},
  {Eisenstein}, {Evans}, {Fiorucci}, {Fulbright}, {Gerhard}, {Jauregi}, {Kelz},
  {Mijovi{\'c}}, {Minchev}, {Parmentier}, {Pe{\~n}arrubia}, {Quillen}, {Read},
  {Ruchti}, {Scholz}, {Siviero}, {Smith}, {Sordo}, {Veltz}, {Vidrih}, {von
  Berlepsch}, {Boyle}, \& {Schilbach}}]{2006AJ....132.1645S}
{Steinmetz}, M., {Zwitter}, T., {Siebert}, A., {et~al.} 2006, \aj, 132, 1645

\bibitem[{{Tojeiro} {et~al.}(2007){Tojeiro}, {Heavens}, {Jimenez}, \&
  {Panter}}]{2007MNRAS.381.1252T}
{Tojeiro}, R., {Heavens}, A.~F., {Jimenez}, R., \& {Panter}, B. 2007, \mnras,
  381, 1252

\bibitem[{{Trujillo} {et~al.}(2001){Trujillo}, {Aguerri}, {Cepa}, \&
  {Guti{\'e}rrez}}]{2001MNRAS.328..977T}
{Trujillo}, I., {Aguerri}, J.~A.~L., {Cepa}, J., \& {Guti{\'e}rrez}, C.~M.
  2001, \mnras, 328, 977

\bibitem[{{Valdes} {et~al.}(2004){Valdes}, {Gupta}, {Rose}, {Singh}, \&
  {Bell}}]{2004ApJS..152..251V}
{Valdes}, F., {Gupta}, R., {Rose}, J.~A., {Singh}, H.~P., \& {Bell}, D.~J.
  2004, \apjs, 152, 251

\bibitem[{{van den Bosch}(2001)}]{2001MNRAS.327.1334V}
{van den Bosch}, F.~C. 2001, \mnras, 327, 1334

\bibitem[{{van der Marel} \& {Franx}(1993)}]{1993ApJ...407..525V}
{van der Marel}, R.~P. \& {Franx}, M. 1993, \apj, 407, 525

\bibitem[{{Vazdekis} {et~al.}(1996){Vazdekis}, {Casuso}, {Peletier}, \&
  {Beckman}}]{1996ApJS..106..307V}
{Vazdekis}, A., {Casuso}, E., {Peletier}, R.~F., \& {Beckman}, J.~E. 1996,
  \apjs, 106, 307

\bibitem[{{Vazdekis} {et~al.}(2010){Vazdekis}, {S{\'a}nchez-Bl{\'a}zquez},
  {Falc{\'o}n-Barroso}, {Cenarro}, {Beasley}, {Cardiel}, {Gorgas}, \&
  {Peletier}}]{2010MNRAS.404.1639V}
{Vazdekis}, A., {S{\'a}nchez-Bl{\'a}zquez}, P., {Falc{\'o}n-Barroso}, J.,
  {et~al.} 2010, \mnras, 404, 1639

\bibitem[{{Walcher} {et~al.}(2014){Walcher}, {Wisotzki}, {Bekerait{\'e}},
  {Husemann}, {Iglesias-P{\'a}ramo}, {Backsmann}, {Barrera Ballesteros},
  {Catal{\'a}n-Torrecilla}, {Cortijo}, {del Olmo}, {Garcia Lorenzo},
  {Falc{\'o}n-Barroso}, {Jilkova}, {Kalinova}, {Mast}, {Marino},
  {M{\'e}ndez-Abreu}, {Pasquali}, {S{\'a}nchez}, {Trager}, {Zibetti},
  {Aguerri}, {Alves}, {Bland-Hawthorn}, {Boselli}, {Castillo Morales}, {Cid
  Fernandes}, {Flores}, {Galbany}, {Gallazzi}, {Garc{\'{\i}}a-Benito}, {Gil de
  Paz}, {Gonz{\'a}lez-Delgado}, {Jahnke}, {Jungwiert}, {Kehrig}, {Lyubenova},
  {M{\'a}rquez Perez}, {Masegosa}, {Monreal Ibero}, {P{\'e}rez}, {Quirrenbach},
  {Rosales-Ortega}, {Roth}, {Sanchez-Blazquez}, {Spekkens}, {Tundo}, {van de
  Ven}, {Verheijen}, {Vilchez}, \& {Ziegler}}]{2014A&A...569A...1W}
{Walcher}, C.~J., {Wisotzki}, L., {Bekerait{\'e}}, S., {et~al.} 2014, \aap,
  569, A1

\bibitem[{{Worthey}(1994)}]{1994ApJS...95..107W}
{Worthey}, G. 1994, \apjs, 95, 107

\bibitem[{{Worthey} \& {Ottaviani}(1997)}]{1997ApJS..111..377W}
{Worthey}, G. \& {Ottaviani}, D.~L. 1997, \apjs, 111, 377

\bibitem[{{Yoachim} {et~al.}(2012){Yoachim}, {Ro{\v s}kar}, \&
  {Debattista}}]{2012ApJ...752...97Y}
{Yoachim}, P., {Ro{\v s}kar}, R., \& {Debattista}, V.~P. 2012, \apj, 752, 97

\bibitem[{{York} {et~al.}(2000){York}, {Adelman}, {Anderson}, {Anderson},
  {Annis}, {Bahcall}, {Bakken}, {Barkhouser}, {Bastian}, {Berman}, {Boroski},
  {Bracker}, {Briegel}, {Briggs}, {Brinkmann}, {Brunner}, {Burles}, {Carey},
  {Carr}, {Castander}, {Chen}, {Colestock}, {Connolly}, {Crocker}, {Csabai},
  {Czarapata}, {Davis}, {Doi}, {Dombeck}, {Eisenstein}, {Ellman}, {Elms},
  {Evans}, {Fan}, {Federwitz}, {Fiscelli}, {Friedman}, {Frieman}, {Fukugita},
  {Gillespie}, {Gunn}, {Gurbani}, {de Haas}, {Haldeman}, {Harris}, {Hayes},
  {Heckman}, {Hennessy}, {Hindsley}, {Holm}, {Holmgren}, {Huang}, {Hull},
  {Husby}, {Ichikawa}, {Ichikawa}, {Ivezi{\'c}}, {Kent}, {Kim}, {Kinney},
  {Klaene}, {Kleinman}, {Kleinman}, {Knapp}, {Korienek}, {Kron}, {Kunszt},
  {Lamb}, {Lee}, {Leger}, {Limmongkol}, {Lindenmeyer}, {Long}, {Loomis},
  {Loveday}, {Lucinio}, {Lupton}, {MacKinnon}, {Mannery}, {Mantsch}, {Margon},
  {McGehee}, {McKay}, {Meiksin}, {Merelli}, {Monet}, {Munn}, {Narayanan},
  {Nash}, {Neilsen}, {Neswold}, {Newberg}, {Nichol}, {Nicinski}, {Nonino},
  {Okada}, {Okamura}, {Ostriker}, {Owen}, {Pauls}, {Peoples}, {Peterson},
  {Petravick}, {Pier}, {Pope}, {Pordes}, {Prosapio}, {Rechenmacher}, {Quinn},
  {Richards}, {Richmond}, {Rivetta}, {Rockosi}, {Ruthmansdorfer}, {Sandford},
  {Schlegel}, {Schneider}, {Sekiguchi}, {Sergey}, {Shimasaku}, {Siegmund},
  {Smee}, {Smith}, {Snedden}, {Stone}, {Stoughton}, {Strauss}, {Stubbs},
  {SubbaRao}, {Szalay}, {Szapudi}, {Szokoly}, {Thakar}, {Tremonti}, {Tucker},
  {Uomoto}, {Vanden Berk}, {Vogeley}, {Waddell}, {Wang}, {Watanabe},
  {Weinberg}, {Yanny}, {Yasuda}, \& {SDSS Collaboration}}]{2000AJ....120.1579Y}
{York}, D.~G., {Adelman}, J., {Anderson}, Jr., J.~E., {et~al.} 2000, \aj, 120,
  1579

\bibitem[{{Younger} {et~al.}(2007){Younger}, {Cox}, {Seth}, \&
  {Hernquist}}]{2007ApJ...670..269Y}
{Younger}, J.~D., {Cox}, T.~J., {Seth}, A.~C., \& {Hernquist}, L. 2007, \apj,
  670, 269

\bibitem[{{Zhang} {et~al.}(2012){Zhang}, {Hunter}, {Elmegreen}, {Gao}, \&
  {Schruba}}]{2012AJ....143...47Z}
{Zhang}, H.-X., {Hunter}, D.~A., {Elmegreen}, B.~G., {Gao}, Y., \& {Schruba},
  A. 2012, \aj, 143, 47

\bibitem[{{Zheng} {et~al.}(2015){Zheng}, {Thilker}, {Heckman}, {Meurer},
  {Burgett}, {Chambers}, {Huber}, {Kaiser}, {Magnier}, {Metcalfe}, {Price},
  {Tonry}, {Wainscoat}, \& {Waters}}]{2015ApJ...800..120Z}
{Zheng}, Z., {Thilker}, D.~A., {Heckman}, T.~M., {et~al.} 2015, \apj, 800, 120

\bibitem[{{Zibetti} {et~al.}(2017){Zibetti}, {Gallazzi}, {Ascasibar},
  {Charlot}, {Galbany}, {Garc{\'{\i}}a Benito}, {Kehrig}, {de
  Lorenzo-C{\'a}ceres}, {Lyubenova}, {Marino}, {M{\'a}rquez}, {S{\'a}nchez},
  {van de Ven}, {Walcher}, \& {Wisotzki}}]{2017MNRAS.468.1902Z}
{Zibetti}, S., {Gallazzi}, A.~R., {Ascasibar}, Y., {et~al.} 2017, \mnras, 468,
  1902

\end{thebibliography}
\bibliographystyle{aa} 

\appendix

\section{SB profile classification}
\label{II_SB_prof_class}

The proper characterization of the light distribution in the galaxy sample is an essential step to analyse the stellar content for different SB profiles. Table~\ref{II_tab_sample_SB_I} summarise the information from the 2D light decomposition presented in this work (see Sect.~\ref{II_jairo}). We find that type I (61.7 \%) and II (32.2 \%) SB profiles are the most frequent ones in the galaxies under analysis with only 13 galaxies (6.1 \%) displaying an upbending profile. Type II galaxies exhibit the most extended inner discs (h$_{\rm in}$~=~8.92~$\pm$~3.80 kpc), followed by type I (4.88~$\pm$~1.15 kpc) and type III (2.51~$\pm$~0.36 kpc) discs. The breaks are located at larger galactocentric distances for type II than for type III galaxies (12.31~$\pm$~3.25 vs. 8.50~$\pm$~1.40 kpc), although in terms of h$_{\rm in}$ this trend is reversed (1.92~$\pm$~0.68 vs. 3.51~$\pm$~0.54 h$_{\rm in}$). There seems to be a tendency with morphology in the inner disc-scalelength for type I galaxies in which later types show less extended discs than early types. However, no further correlations are found with morphology.

\begin{table*}
\small
\centering
\begin{tabular}{lcccc@{\hspace{8mm}}cccccc}
\hline
               &    N    & h$_{\rm in}$(kpc) & R$_{\rm break}$(kpc) &  R$_{\rm break}$/h$_{\rm in}$ &  &  N & h$_{\rm in}$(kpc) &  R$_{\rm break}$(kpc) &  R$_{\rm break}$/h$_{\rm in}$ \\
\hline
  \multirow{4}{*}{{\bf Type I}} & \multirow{4}{*}{132 (61.7 $\%$)} & \multirow{4}{*}{4.88~$\pm$~1.15} & \multirow{4}{*}{--} & \multirow{4}{*}{--} & Sa & 22 (10.3) & 5.22~$\pm$~1.41 & -- & --\\
                                & & &  & & Sb & 47 (22.0) & 5.22$\pm$ 0.97 & -- & --\\
                                & & & & & Sc & 52 (24.3) & 4.66$\pm$ 1.15 &-- & --\\
                                & & & & & Sd & 11 (5.1) & 3.75$\pm$ 1.21 & --& --\\
\hline
  \multirow{4}{*}{{\bf Type II}} & \multirow{4}{*}{69 (32.2$\%$)}  & \multirow{4}{*}{8.92$\pm$ 3.80}& \multirow{4}{*}{12.31$\pm$3.25} & \multirow{4}{*}{1.92~$\pm$~0.68}&  Sa & 8 (3.7) & 8.22$\pm$ 2.55 & 11.09$\pm$2.92 & 2.26$\pm$ 1.33\\
                                & & & & & Sb & 19 (8.9) & 12.15$\pm$ 2.61 & 11.77$\pm$ 2.89 & 1.40$\pm$ 0.41\\
                                & & & & & Sc & 31 (14.5) & 8.78$\pm$ 3.15 & 13.86$\pm$3.45 & 2.04$\pm$ 0.57 \\
                                & & & & & Sd & 11 (5.1) & 4.26$\pm$ 0.85 & 9.77 $\pm$2.90 & 2.20$\pm$ 0.37\\
\hline
  \multirow{4}{*}{{\bf Type III}} & \multirow{4}{*}{13 (6.1$\%$)} & \multirow{4}{*}{2.51$\pm$0.36}& \multirow{4}{*}{8.50$\pm$1.40}&\multirow{4}{*}{3.51~$\pm$~0.54}&  Sa & 5 (2.3) & 2.98~$\pm$~0.25  & 9.65~$\pm$~1.77 & 7.17~$\pm$~0.45 \\
                                & & &  & & Sb & 2 (0.9) & 2.55$\pm$ 0.09  & 7.41$\pm$0.61 & 2.89$\pm$ 0.14\\
                                & & & & & Sc & 4 (2.0) & 1.89$\pm$ 0.29 & 8.58$\pm$1.05 & 4.67$\pm$ 0.40\\
                                & & & & & Sd & 2 (0.9) & 2.51~$\pm$~0.37  & 6.55~$\pm$~0.78 & 2.62~$\pm$~0.09 \\
\hline
\end{tabular}\vspace{2mm}
\caption{Surface Brightness profile parameters for the sample of galaxies from the 2D decomposition using GASP2D (see Sect.~\ref{II_jairo} for further details).}
\label{II_tab_sample_SB_I}
\end{table*}

Although it is not the main scope of this paper, for the sake of completeness, in Table~\ref{II_tab_sample_SB_II} we compare our results with other works focused on the 1D light distribution up to the outer discs using visible light, i.e. \citet[][hereafter PT06]{2006A&A...454..759P}, \citet[][hereafter E08]{2008AJ....135...20E}, \citet[][hereafter G11]{2011AJ....142..145G}, and \citet[][hereafter M16]{2016A&A...585A..47M}. We should warn the reader that, by no means, this comparison is meant to be a complete one: there are many more works in the literature to be compare with. We must also highlight the different selection criteria adopted to define each sample. PT06 was focused on analysing the light distribution for late type galaxies (98 galaxies with 2.99~< T~< 8.49), while E08 and G11 were focused on S0--Sb galaxies (66 barred and 47 unbarred galaxies, respectively). The sample analysed in M16 should, in principle, present more similarities with the one analysed in this work as both were drawn from the CALIFA sample (with some overlapping between both samples).

The comparison among these works (including the one presented here) shows some differences. Especially striking is the discrepancy in the frequency of profile types obtained by these studies. While this work suggests that single exponential (type I) profiles are the most frequent type of SB profiles found in our galaxies, PT06, E08, and M16 suggest that galaxies mostly display downbending profiles, and upbending SB distributions are most frequently found in G11. However, we find a lack of type III galaxies with the 2D decomposition method. The values for the inner disc-scalelengths in all the works suggest that type II galaxies have larger values than type I and III galaxies (in that order), with the exception of G11 for which the order would change to type I, II, and III with decreasing inner disc-scalelength values. However, while all the works in 1D found relatively consistent values for this morphological parameter (within errors), the 2D analysis presented here arises larger inner disc-scalelengths for type I and II galaxies than the rest of the works. Regarding the break radius, although in physical units different works found different values (both for type II and type III systems), those found for the R$_{\rm break}$/h$_{\rm in}$ for different SB types are fairly consistent among works, with values around 2 for type II galaxies and around 4 for type III systems. All the works seem to suggest that the position of the break in units of h$_{\rm in}$ is larger for type III galaxies than for downbending profiles while in terms of physical units there is no agreement between them. The tendency found with this 2D analysis for type I galaxies with later types displaying lower values of h$_{\rm in}$ is also found when comparing the PT06 results (later types) with the E08 or G11 ones (earlier types).

Before drawing any conclusions, we must bear in mind the great discrepancies among the compared works: i) PT06, E08, G11, and M16 follow a 1D analysis of the light distribution while this analysis makes use of a 2D decomposition; in addition, ii) different galaxy samples are analysed in each work; if the samples under analysis are not similar (masses, morphology, etc) then, different results might arise as a consequence of the differences among the analysed samples.

\begin{table*}\centering
\begin{tabular}{lccccccc}
\hline
               &        & This work (2D) & This work (1D) & PT06 &  E08 & G11 & M16 \\
               &        & (1) & (2) & (3) &  (4) & (5) & (6) \\
\hline
  \multirow{4}{*}{{\bf Type I}} & \% & 61.7 & 34.1 & 11 & 27 & 28 & 16.4 \\
                                & h$_{\rm in}$(kpc) & 4.88~$\pm$~1.15 & 4.02~$\pm$~0.79 & 2.8~$\pm$~0.8 & 2.89~$\pm$~0.67 & 5.98~$\pm$~2.70 & 4.53~$\pm$~2.21 \\
                                & R$_{\rm break}$ (kpc) & -- & -- & -- & -- & -- & -- \\
                                & R$_{\rm break}$/h$_{\rm in}$ & -- & -- & -- & -- & -- & -- \\
\hline
  \multirow{4}{*}{{\bf Type II}} & \% & 32.2 & 39.7 & 66 & 42 & 21 & 52.8 \\
                                & h$_{\rm in}$(kpc) & 8.92~$\pm$~3.80 & 6.38~$\pm$~1.91 & 3.8~$\pm$~1.2 / 6.9~$\pm$~2.25 & 6.07~$\pm$~2.50 & 4.46~$\pm$~1.83 & 4.60~$\pm$~0.42 \\
                                & R$_{\rm break}$ (kpc) & 12.31~$\pm$~3.25 & 12.71~$\pm$~2.44 & 9.2~$\pm$~2.4 / 9.5~$\pm$~3.3 & 7.69~$\pm$~1.90 & 7.86~$\pm$~2.35 & 10.51~$\pm$~0.48 \\
                                & R$_{\rm break}$/h$_{\rm in}$ & 1.92~$\pm$~0.68 & 3.57~$\pm$~0.44 & 2.5~$\pm$~0.6 / 1.7~$\pm$~0.4 & 2.07~$\pm$~0.81 & 2.06~$\pm$~0.38 & 2.28~$\pm$~0.64 \\
\hline
  \multirow{4}{*}{{\bf Type III}} & \% & 6.1 & 26.2 & 33 & 24 & 51 & 30.8 \\
                                & h$_{\rm in}$(kpc) & 2.51~$\pm$~0.36 & 3.47~$\pm$~0.91 & 1.9~$\pm$~0.6 & 1.89~$\pm$~0.35 & 2.27~$\pm$~0.51 & 3.12~$\pm$~0.21 \\
                                & R$_{\rm break}$ (kpc) & 8.50~$\pm$~1.40 & 12.15~$\pm$~3.44 & 9.3~$\pm$~3.3 & 8.24~$\pm$~1.65 & 9.44~$\pm$~2.23 & 7.68~$\pm$~0.49\\
                                & R$_{\rm break}$/h$_{\rm in}$ & 3.51~$\pm$~0.54 & 3.57~$\pm$~0.37 & 4.9~$\pm$~0.6 & 4.40~$\pm$~0.41 & 4.24~$\pm$~0.62 & 2.46~$\pm$~0.53\\
\hline
\end{tabular}\vspace{2mm}
\caption{Comparison between the surface brightness profile parameters presented here and other works. (1) Results applying the 2D decomposition using GASP2D (see Sect.~\ref{II_jairo} for further details) to the sample of galaxies under analysis; (2) results applying a more classical 1D approch (see text for details) to the sample of galaxies; (3) results from \citet[][]{2006A&A...454..759P}, for type II galaxies the parameters are divided into classical truncations and breaks associated with the Outer Lindblad Reso--ce (II-CT/II-o.OLR); (4) results from \citet[][]{2008AJ....135...20E}; (5) results from \citet[][]{2011AJ....142..145G}; and (6) results from \citet[][]{2016A&A...585A..47M}.}
\label{II_tab_sample_SB_II}
\end{table*}

In order to check to what extend the decomposition method affects the global morphological parameters for type I, II, and III galaxies we apply a 1D analysis similar to the one performed in other works to the sample of galaxies analysed in this work and compare the results from both methods on the same sample. To perform such exercise we have analysed the 1D SB distributions already computed in Sect.~\ref{II_colours}. We fit these 1D SB profiles with a classical exponential decline \citep[][]{1970ApJ...160..811F} in the case of type I galaxies:
\begin{equation} 
I_{\rm disc}(r)=I_{\rm 0}\, e^{-(\frac{r}{h_{\rm in}})},
\label{eqn:disc} 
\end{equation} 
where I$_{\rm disc}$(r) is the disc intensity as a function of the radius ($r$), $I_{\rm 0}$ is the central surface brightness, and h$_{\rm in}$ is the scale-length of the disc. For type II and III galaxies we use the function defined in E08:
\begin{equation} 
I_{\rm disc}(r)=S\,I_{\rm 0}\, e^{-r/h_{\rm in}} \left[1+e^{\alpha(r-R_{\rm break})}\right]^{\frac{1}{\alpha}\, \left(\frac{1}{h_{\rm in}} - \frac{1}{h_{\rm out}} \right)},
\label{eqn:disc_trunc} 
\end{equation} 
where h$_{\rm out}$ is the outer scale-length of the disc, R$_{\rm break}$ is the position of the break, $\alpha$ is a parametrization of the sharpness of the break, and S is a scaling factor in the form:
\begin{equation} 
S= \left(1+e^{-\alpha R_{\rm break}}\right)^{\frac{1}{\alpha}\, \left(\frac{1}{h_{\rm in}} - \frac{1}{h_{\rm out}} \right)} .
\label{eqn:disc_trunc} 
\end{equation} 

The results of this 1D analysis of the light distribution are shown in column (2) of Table~\ref{II_tab_sample_SB_II}. We find again a discrepancy in the frequency of profile types with this 1D approach displays more similar percentages when compared with other works. Not only the frequency of profile types changes from a 1D to a 2D approach, even analysing the same sample of galaxies, but also the rest of morphological parameters. Different values for the break radii and inner disc-scalelengths are found with no clear pattern. In addition, the comparison of the results from the 1D approach in this sample of galaxies and the results of other works present quite striking differences mainly due to the different samples under analysis.

In summary, the differences among the compared works are mainly due to two aspects. On one hand, the different decomposition methods used (1D vs 2D) must play an important role in the derived structural parameters; however, a 2D analysis of the light distribution is a more realistic procedure, and thus, we decide to follow this approach in this work. On the other hand, the different selection criteria to define each sample and thus, the different morphologies under consideration, might arise different results. It is clear that larger statistical samples and a more rigorous analysis (far from the scope of this present work) are needed to properly assess the variety of SB profiles in nature. These findings are agreement to what shown in \citet[][]{2017A&A...598A..32M}.

\clearpage
\onecolumn

\section{stellar parameter and colour gradients}
\label{sec:appendix1}
In this section we present tables with the derived stellar age and metallicity profile inner gradients as well as colour profile inner gradients for all the galaxies in the sample. Table~\ref{II_tab_ages} gives this information for the stellar age profiles, Table~\ref{II_tab_mets} for the [M/H] profiles, and Table~\ref{II_tab_colours} for the colour ($g-r$, $g-i$, and $r-i$) profiles.

\begin{longtab}
\centering
\LTcapwidth=\textwidth
\begin{longtable}{lrrrr}
\hline\hline\\
 Name & \multicolumn{2}{c}{$\bigtriangledown log(Age)_{L, in}$} & \multicolumn{2}{c}{$\bigtriangledown log(Age)_{M, in}$}  \\    
  & ($dex/h_{in}$) & ($dex/kpc$) &  ($dex/h_{in}$) & ($dex/kpc$)  \\    
(1) & (2) & (3) & (4) & (5)  \\ 
\hline\\
\endfirsthead
\hline\hline\\
 Name & \multicolumn{2}{c}{$\bigtriangledown log(Age)_{L, in}$} & \multicolumn{2}{c}{$\bigtriangledown log(Age)_{M, in}$}  \\    
  & ($dex/h_{in}$) & ($dex/kpc$) &  ($dex/h_{in}$) & ($dex/kpc$)  \\    
(1) & (2) & (3) & (4) & (5)  \\ 
\hline\\
\endhead
\caption{Continued.}\\
\endfoot
\endlastfoot
UGC~00005 & -0.122 $\pm$ 0.008 & -0.019 $\pm$ 0.001 & -0.008 $\pm$ 0.005 & -0.001 $\pm$ 0.001  \\ 
NGC~7819 & -0.098 $\pm$ 0.025 & -0.006 $\pm$ 0.002 & -0.184 $\pm$ 0.016 & -0.012 $\pm$ 0.001  \\ 
IC~1528 & -0.138 $\pm$ 0.007 & -0.033 $\pm$ 0.002 & -0.053 $\pm$ 0.004 & -0.013 $\pm$ 0.001  \\ 
UGC~00036 & -0.207 $\pm$ 0.008 & -0.051 $\pm$ 0.002 & -0.016 $\pm$ 0.005 & -0.004 $\pm$ 0.001  \\ 
NGC~0001 & -0.254 $\pm$ 0.007 & -0.059 $\pm$ 0.002 & -0.001 $\pm$ 0.005 & -0.0 $\pm$ 0.001  \\ 
NGC~0023 & 1.404 $\pm$ 0.116 & 0.041 $\pm$ 0.003 & 0.267 $\pm$ 0.072 & 0.008 $\pm$ 0.002  \\ 
NGC~0036 & -0.414 $\pm$ 0.008 & -0.043 $\pm$ 0.001 & -0.024 $\pm$ 0.005 & -0.003 $\pm$ 0.001  \\ 
UGC~00139 & 0.005 $\pm$ 0.077 & 0.001 $\pm$ 0.024 & -0.04 $\pm$ 0.036 & -0.012 $\pm$ 0.011  \\
\end{longtable}
\end{longtab}
\newpage

\begin{longtab}
\centering
\LTcapwidth=\textwidth
\begin{longtable}{lrrrr}
\hline\hline\\
 Name & \multicolumn{2}{c}{$\bigtriangledown [M/H]_{L, in}$} & \multicolumn{2}{c}{$\bigtriangledown [M/H]_{M, in}$}  \\    
  & ($dex/h_{in}$) & ($dex/kpc$) &  ($dex/h_{in}$) & ($dex/kpc$)  \\    
(1) & (2) & (3) & (4) & (5)  \\ 
\hline\\
\endfirsthead
\hline\hline\\
 Name & \multicolumn{2}{c}{$\bigtriangledown [M/H]_{L, in}$} & \multicolumn{2}{c}{$\bigtriangledown [M/H]_{M, in}$}  \\    
  & ($dex/h_{in}$) & ($dex/kpc$) &  ($dex/h_{in}$) & ($dex/kpc$)  \\    
(1) & (2) & (3) & (4) & (5)  \\ 
\hline\\
\endhead
\caption{Continued.}\\
\endfoot
\endlastfoot
UGC~00005 & -0.176 $\pm$ 0.009 & -0.028 $\pm$ 0.001 & 0.007 $\pm$ 0.011 & 0.001 $\pm$ 0.002  \\ 
NGC~7819 & -0.124 $\pm$ 0.027 & -0.008 $\pm$ 0.002 & -0.752 $\pm$ 0.037 & -0.049 $\pm$ 0.002  \\ 
IC~1528 & -0.017 $\pm$ 0.009 & -0.004 $\pm$ 0.002 & -0.15 $\pm$ 0.012 & -0.036 $\pm$ 0.003  \\ 
UGC~00036 & -0.043 $\pm$ 0.006 & -0.01 $\pm$ 0.001 & 0.046 $\pm$ 0.007 & 0.011 $\pm$ 0.002  \\ 
NGC~0001 & -0.071 $\pm$ 0.01 & -0.016 $\pm$ 0.002 & 0.018 $\pm$ 0.013 & 0.004 $\pm$ 0.003  \\ 
NGC~0023 & 1.887 $\pm$ 0.105 & 0.055 $\pm$ 0.003 & 0.026 $\pm$ 0.121 & 0.001 $\pm$ 0.004  \\ 
NGC~0036 & -0.077 $\pm$ 0.011 & -0.008 $\pm$ 0.001 & 0.101 $\pm$ 0.014 & 0.011 $\pm$ 0.001  \\ 
UGC~00139 & -0.02 $\pm$ 0.033 & -0.006 $\pm$ 0.01 & -0.525 $\pm$ 0.05 & -0.161 $\pm$ 0.015  \\
\end{longtable}
\end{longtab}
\newpage

\begin{longtab}
\small
\centering
\LTcapwidth=\textwidth
\begin{longtable}{lrrrrrr}
\hline\hline\\
 Name & \multicolumn{2}{c}{$\bigtriangledown (g-r)_{in}$} & \multicolumn{2}{c}{$\bigtriangledown (g-i)_{in}$} & \multicolumn{2}{c}{$\bigtriangledown (r-i)_{in}$}  \\    
  & ($dex/h_{in}$) & ($dex/kpc$) &  ($dex/h_{in}$) & ($dex/kpc$) & ($dex/h_{in}$) & ($dex/kpc$) \\    
(1) & (2) & (3) & (4) & (5) & (6) & (7) \\ 
\hline\\
\endfirsthead
\hline\hline\\
 Name & \multicolumn{2}{c}{$\bigtriangledown (g-r)_{in}$} & \multicolumn{2}{c}{$\bigtriangledown (g-i)_{in}$} & \multicolumn{2}{c}{$\bigtriangledown (r-i)_{in}$}  \\    
 & ($dex/h_{in}$) & ($dex/kpc$) &  ($dex/h_{in}$) & ($dex/kpc$) & ($dex/h_{in}$) & ($dex/kpc$) \\    
(1) & (2) & (3) & (4) & (5) & (6) & (7) \\ 
\hline\\
\endhead
\caption{Continued.}\\
\endfoot
\endlastfoot
UGC~00005 & -0.13 $\pm$ 0.18 & -0.02 $\pm$ 0.03 & -0.20 $\pm$ 0.18 & -0.03 $\pm$ 0.03 & -0.07 $\pm$ 0.18 & -0.01 $\pm$ 0.03  \\ 
NGC~7819 & -0.33 $\pm$ 0.36 & -0.02 $\pm$ 0.02 & -0.49 $\pm$ 0.36 & -0.03 $\pm$ 0.02 & -0.16 $\pm$ 0.36 & -0.01 $\pm$ 0.02  \\ 
IC~1528 & -0.10 $\pm$ 0.12 & -0.02 $\pm$ 0.03 & -0.12 $\pm$ 0.12 & -0.03 $\pm$ 0.03 & -0.02 $\pm$ 0.12 & -0.01 $\pm$ 0.03  \\ 
UGC~00036 & -0.04 $\pm$ 0.16 & -0.01 $\pm$ 0.04 & -0.05 $\pm$ 0.16 & -0.01 $\pm$ 0.04 & -0.00 $\pm$ 0.16 & -0.00 $\pm$ 0.04  \\ 
NGC~0001 & -0.07 $\pm$ 0.13 & -0.02 $\pm$ 0.03 & -0.08 $\pm$ 0.13 & -0.02 $\pm$ 0.03 & -0.02 $\pm$ 0.13 & -0.00 $\pm$ 0.03  \\ 
NGC~0023 & -0.33 $\pm$ 0.78 & -0.01 $\pm$ 0.02 & -0.32 $\pm$ 0.78 & -0.01 $\pm$ 0.02 & 0.02 $\pm$ 0.78 & 0.00 $\pm$ 0.02  \\ 
NGC~0036 & -0.16 $\pm$ 0.17 & -0.02 $\pm$ 0.02 & -0.33 $\pm$ 0.17 & -0.03 $\pm$ 0.02 & -0.17 $\pm$ 0.17 & -0.02 $\pm$ 0.02  \\ 
UGC~00139 & -0.19 $\pm$ 0.21 & -0.06 $\pm$ 0.06 & -0.26 $\pm$ 0.21 & -0.08 $\pm$ 0.06 & -0.07 $\pm$ 0.21 & -0.02 $\pm$ 0.06  \\
\end{longtable}
\end{longtab}
\newpage

\clearpage
\twocolumn

\end{document}